# The Y-stringlike behaviour of a static hadron at finite T.

Ahmed S. Bakry,[1,*] Xurong Chen,[1] and Peng-Ming Zhang[1,2]

[1] Institute of Modern Physics, Chinese Academy of Sciences, Lanzhou, China
[2] State Key Laboratory of Theoretical Physics, Institute of Theoretical Physics, Chinese Academy of Sciences, Beijing 100190, China

We look into the signatures of the effective Y-bosonic strings in the gluonic profile due to a system of three static quarks on the lattice. The gluonic distribution is calculated in pure Yang-Mills lattice gauge theory at finite temperature with Polyakov loops operators. The analysis of the action density unveils a filled-$\Delta$ distribution. However, we found that these $\Delta$-shaped action density profiles are structured from three Y-shaped Gaussian-like flux-tubes. The length of the revealed Y-flux system increases with the increasing of the color source separation and reaches maximum near the deconfinement point. The lattice data for the mean-square width of the gluonic action density have been compared to the corresponding width calculated based on the string model at finite temperature. We assume Y-string configuration with minimal length. The growth pattern of the action density of the gluonic field fits well to junction fluctuations of the Y-baryonic string model for large quark separation at the considered temperatures.

## I. INTRODUCTION

The confinement of quarks into mesons and hadrons is an outstanding feature of quantum chromodynamics (QCD). Computer simulations have revealed that quark confinement is a property of the gluonic sector and is common for non-Abelian gauge models [1–4]. The static quark–antiquark potential is linearly rising [5] with interquark separation. The origin of the linearly rising confining potential has been conjectured, in a string-like flux tube model, to be due to the formation of a stringlike flux tube [6–12] between the two quark color sources.

The IR dynamics of the gluonic sector in the meson have shown gross features to be described in a string picture [7–12]. The squeezed flux-tube, by virtue of the surrounding superconductiong medium [2, 3, 13], is conjectured to vibrate, after roughening, like a massless string. The profile of the vibrating flux tube can be unraveled in numerical lattice simulations by correlating the field strength of the QCD vacuum to the constructed quark states [14–17]. The large distance properties of the energy distribution in the meson have been a subject of many lattice simulations targeting the properties of the flux tubes at both zero and finite temperatures. The string picture's main measurable universal consequences of the Lüscher subleading correction to the $Q\overline{Q}$ potential [12] and flux-tube logarithmic growth law [18] have been verified with the lattice data at large distances [4, 9, 14, 19, 19–22].

Nevertheless, the study of the stringlike behavior of flux tubes in multi-quark systems [23] seems to be less visited on the lattice. The calculations are prone to practical difficulties associated with both systematic and statistical uncertainties. The signal in the baryon is noisy [24], and the form of field distribution of the physically interesting ground state seems less obvious [24, 25] than the mesonic case. There have been systematic difficulties to unravel an unbiased form of the action density distribution within the framework of the overlap formalism, i.e. employing Wilson loops as a quark source operator [24–26]. The gluonic wave function is reflected in the form of the field distribution [24–26]. This has presented hitherto a challenge to directly scrutinize unambiguously the baryonic strings on the lattice.

Our recent lattice Monte-Carlo simulations, however, have unraveled the distribution of the gluonic gauge field inside a static baryon at finite temperature [27]. The action density correlation with three Polyakov loops, representing an infinitely heavy quark state in pure SU(3) Yang-Mills theory, has displayed a filled $\Delta$-shaped profile [27, 28]. This filled $\Delta$-shaped arrangement interestingly persists to large interquark separations [27]. This varies the action density profile obtained using Wilson loops as a quark operator [24–26]. Wilson's loop operator displays three distinctive Y-shaped flux tubes forms in the interquark space at large distances [26]. The returned gluonic pattern of the baryon with three Polyakov loops displayed a system rich of features [27, 28] taking into account the role played by the temperature as well. The associated signatures of the string configuration relevant to this quark system [29, 30] is indeed an interesting topic that remains to be addressed in detail.

Static color charges corresponding to a multi-quark system may induce an intricate stringy system in the QCD vacuum including the formation of multi-junction systems [23, 31]. In the baryon, the picture would simplify to the Y-string configuration which is expected to be a staple configuration in the IR region of the baryon [31]. This system accounts for three strings originating from a node to the 3-quarks [32]. The modeling of this system would entail utilizing collective co-ordinates referring to the junction location. Now we have a theoretical development relevant to the effective strings effects in this model [29, 30, 33]. The calculations of the Casimir energy have indicated a Lüscher-like subleading term for the Y (3Q) potential [29, 34]. There is also a discussion from a gauge-string perspective in Ref [35].

* abakry@impcas.ac.cn

In fact, a look at the signatures of the baryonic strings in lattice data simultaneously scrutinizes the subleading properties of the quark potential as well as the configuration of the strings that join the constituent quarks; or in other words, the leading term of the 3Q potential. The leading properties of the 3Q potential have been extensively studied on the lattice [36–39] employing various techniques. Recent lattice QCD findings regarding the 3-quark potential have shown that the confining potential admits two possible models depending on the inter-quark separation distances [36–40]. The so-called $\Delta$ parametrization for small quark separation distances of $R < 0.7$ fm and the Y-ansatz for $0.7 < R < 1.5$ fm [38]. In fact, the distances over which the $\Delta$ and $Y$ ansatz parametrization interpolates has been a controversial for a long period of time [36–39] due to the small value in the difference between both ansatz (of the order of 15 %) and the degree of accuracy of the data.

A direct test of the baryonic subleading string signatures in the lattice data for the potential of three static quarks at zero temperature has been reported in Ref. [34]. The numerical measurements of a 3-state Potts gauge model is consistent with the predicted Lüscher-like corrections and the formation of a system of three flux tubes that meet at a junction when the separation between any two quarks is large [34].

In general, the lattice data are in favor of the expected Y-string configuration as the most relevant picture in the IR region of the baryon [41–44]. The quantum delocalization of this string system from its classical configuration results in a mean-square width of the flux distribution. Recently, the study of the dynamics of the junction of the three Y-shaped baryonic flux tubes has shown that the asymptotic behavior of the effective width of the junction grows logarithmically [30] with the distance between the sources. This result [30] is evaluated for equilateral triangular quark configurations at zero temperature. The result resembles the logarithmic growth property of the mesonic flux tubes [4, 9, 14, 19, 19–22] on the lattice.

The feasibility of reproducing lattice data corresponding to the gluonic pattern in a three-quark system at two temperatures below the deconfinement point presents a tempting opportunity to directly look into the baryonic strings in the properties of the QCD vacuum on a first principles basis. In addition to this, one would like to ascertain the interesting long distance $\Delta$-shaped flux arrangement as a consequences of the stringy aspects of gluonic configurations of a static baryon.

In this work, we study the width profile of the junction due to a Y-string model [29, 30, 34] at finite temperature. The width pattern of the gluonic action density resulting from different three sets of 3Q configurations is investigated versus a variety of Y-shaped 3 string configurations obtained by varying the position of the junction. The fit analysis is performed at two temperatures corresponding to $T = 0.8\,T_c$, and $T = 0.9\,T_c$ which correspond to a temperature close to the end of the QCD plateau and to the deconfinement point, respectively.

The present paper is sectioned as follows: In the first Section II we describe the details of the simulations and noise reduction techniques. The baryonic string model at finite temperature is discussed in Section III. In Section IV, we show the properties of the density distribution and compare the profile of the mean width of the junction fluctuations for various string configuration to the width of gluonic action density of the corresponding quark configurations at two temperatures. In the last Section V, the conclusion is provided.

## II. MEASUREMENTS AND ULTRAVIOLET FILTERING

### A. Color field measurements

The heavy baryonic state is constructed by means of Polyakov loop correlators,

$$\mathcal{P}_{3Q} = \langle P(\vec{r}_1) P(\vec{r}_2) P(\vec{r}_3) \rangle,$$

where the color-averaged Polyakov loop is given by

$$P(\vec{r}_i) = \frac{1}{3} \mathrm{Tr} \left[ \prod_{n_t=1}^{N_t} U_{\mu=4}(\vec{r}_i, n_t) \right],$$

and the vectors $\vec{r}_i$ define the positions of the quarks. The measurements that characterize the color field are taken by a gauge-invariant action density operator $S(\vec{\rho}, t)$ at spatial coordinate $\vec{\rho}$ of the three dimensional torus corresponding to an Euclidean time $t$. The measurements are repeated for each time slice and then averaged,

$$S(\vec{\rho}) = \frac{1}{N_t} \sum_{n_t=1}^{N_t} S(\vec{\rho}, t). \qquad (1)$$

A dimensionless scalar field that characterizes the gluonic field can be defined as

$$\mathcal{C}_B(\vec{\rho}, \vec{r}_1, \vec{r}_2, \vec{r}_3) = \frac{\langle \mathcal{P}_{3Q}(\vec{r}_1, \vec{r}_2, \vec{r}_3)\, S(\vec{\rho}) \rangle}{\langle \mathcal{P}_{3Q}(\vec{r}_1, \vec{r}_2, \vec{r}_3) \rangle \, \langle S(\vec{\rho}) \rangle}, \qquad (2)$$

where $< ...... >$ denotes averaging over configurations and lattice symmetries, and the vector $\vec{\rho}$ refers to the spatial position of the flux probe with respect to some origin. Due to cluster decomposition of the operators, $C$ should approach a value $C \simeq 1$ away from the interquark space. For noise reduction, we make use of translational invariance by computing the correlation on every node of the lattice, averaging the results over the volume of the three-dimensional torus, in addition to the averaging the action measurements taken at each time slice in Eq. (1).

The gauge configurations were generated using the standard Wilson gauge action. The two lattices employed in this investigation are of a typical spatial size of $3.6^3\mathrm{fm}^3$. Performing the simulations on large enough lattice sizes would be beneficial to gain high statistics in a gauge-independent manner and also minimizing the mirror effects and correlations across the boundaries as a by-product [26, 45].

The SU(3) gluonic gauge configurations has been generated employing a pseudo-heatbath algorithm [46, 47] updating the corresponding three SU(2) subgroup elements [48]. Each update step consists of one heatbath and 5 micro-canonical reflections. We chose to perform our analysis with lattices as fine as $a = 0.1$ fm by adopting a coupling of value $\beta = 6.00$, with temporal extents of $\mathrm{N_t} = 8$, and $\mathrm{N_t} = 10$ slices, which correspond to temperatures $\mathrm{T} \simeq 0.9\,\mathrm{T_c}$, and $\mathrm{T} \simeq 0.8\,\mathrm{T_c}$, respectively.

We perform a set of measurements $n_{\mathrm{sub}} = 20$ separated by 70 sweeps of updates. Each set of measurements is taken following 2000 updating sweeps. These sub-measurements are binned together in evaluating Eq. 2. The total measurements taken on 500 bins. In this investigation, we have taken 10,000 measurement at each temperature. The measurements are taken on hierarchically generated configurations.

### B. Ultraviolet filtering

An ultraviolet filtering (UV) step precedes our measurements of the action density distribution throughout the lattice. The UV-filtering of the gauge configurations suppresses the short distance quantum fluctuations of the vacuum and is beneficial in attaining a good signal to noise ratio in the correlations Eq 2. This involves a local action reduction by smearing the gauge links of the whole 4 dimensional lattice.

Smoothing the gauge fields complements our use of lattice symmetries to gain noise reduction in our measurement setup. We have shown previously [49] through a systematic study of the effects of smearing on the flux tube-width profile that the effective string physics in the heavy meson are independent of the UV fluctuations at large source separations. In addition to this, we have found the gauge field smoothing, at the intermediate source separation distance at high temperatures.(where the free string picture is known to poorly describe the flux tube width profile) is improving the behavior of the lattice data in favor of the predictions of the free string model.

Variant to [16] where the Cabbibo-Marinari cooling has been employed, we have chosen to smooth the gauge field by an over-improved stout-link smearing algorithm [50]. In standard stout-link smearing [51], all the links are simultaneously updated. Each sweep of the update consists of a replacement of all the links by the smeared links

$$\tilde{U}_\mu(x) = \exp(iQ_\mu(x))\,U_\mu(x)\,, \tag{3}$$

with

$$Q_\mu(x) = \frac{i}{2}(\Omega^\dagger_\mu(x) - \Omega_\mu(x)) - \frac{i}{6}\mathrm{tr}(\Omega^\dagger_\mu(\mathrm{x}) - \Omega_\mu(\mathrm{x}))\,,$$

and

$$\Omega_\mu(x) = \left(\sum_{\nu\neq\mu}\rho_{\mu\nu}\Sigma^\dagger_{\mu\nu}(x)\right)U^\dagger_\mu(x)\,,$$

where $\Sigma_{\mu\nu}(x)$ denotes the sum of the two staples touching $U_\mu(x)$ which reside in the $\mu-\nu$ plane.

The scheme of over-improvement requires $\Sigma_{\mu\nu}(x)$ to be replaced by a combination of plaquette and rectangular staples. This ratio is tuned by the parameter $\epsilon$ [50]. In the following we use a value of $\epsilon = -0.25$, with $\rho_\mu = \rho = 0.06$. We note that for a value of $\rho = 0.06$ in the over-improved stout-link algorithm is roughly equivalent in terms of UV filtering to the standard stout-link smearing algorithm with the same $\rho = 0.06$.

In a baryonic analysis a contingent inconsistency between the UV-filtered lattice data and the string model in the intermediate source separations may, thus, be indicating a prevalence or overlapping of string configurations other than Y-configuration at these distance. We have considered a typical number of 4-D smearing sweeps corresponding to $n_{sw} = 60, 80$ of stout-link smearing.

## III. BARYONIC STRING MODEL

In the dual superconductor model of the QCD vacuum, the QCD vacuum squeezes the color fields into a confining string dual to the Abrikosov line by the dual Meissner effect [2, 3]. With this intuitive picture, an idealized string-like system of flux tubes [12] transmitting the strongly interacting forces between the color sources was proposed previously [6, 12]. The formation of string-like defects is not a peculiar property of the QCD flux tubes, and is realized in many physical phenomena such as vortices in superfluids [52], flux tubes in superconductors [53], vortices in Bose Einstein condensates [54], Nielsen-Olesen vortices of field theory [55], and cosmic strings [56]. The physical parameters of each of these models fix the properties of this stringlike object.

However, quantum mechanical effects become relevant in certain phases of the model, giving rise to interesting measurable effects. To find a consistent quantum description within the quantization scheme used in bosonic string theories, we encounter the difficulty that this is only possible in 26 dimensions.

An effective description with strings [57]in four dimensions predicts logarithmic growth and a long non-Coulombic term to the quark anti-quark potential well known as the Lüscher term. These predictions have been verified in confining gauge theories on distance scales larger than the intrinsic thickness of the flux tube $1/T_c$ [19] at zero temperature in the so-called rough phase of lattice gauge theories (LGT) [9, 14, 19, 58, 59].

The roughening transition signifies the substantial change in the behavior of the profile of the flux-tube between a quark–antiquark pair from the constant width into logarithmic increase [18] by virtue of the strongly fluctuating underlying string. The transition proceeds with the decrease of the coupling constants and, in this phase, the flux-tube admits a collective co-ordinate description.

At sufficiently high temperature, the equations of motion of a Nambu-Goto type bosonic strings are indicating a linear growth in the tube's width if solved [60] for the width of the action density at the middle plane between two quarks. This prediction has been also verified in LGT by studying configurations with a static quark and antiquark pair [21, 49, 60–63] at large separations and near the deconfinement point. The string model assumptions of the effective description of the tube with a collective coordinate referring to the underlying thin-string seems to be working at high temperature.

The above discussion concerning the validity of the model assumption at high temperatures and near the deconfinement point is of particular relevance especially when discussing a Y-shaped baryonic string model [29, 30, 34] to scrutinize the large distance $\Delta$ baryonic flux arrangement [27, 28].

It is widely accepted that the Y-shaped string is the relevant picture of the baryonic flux tubes to the IR region of the non-Abelian gauge and amounts to three squeezed flux tubes that meet at a junction. Indeed, it can be derived from the strong coupling approximation and is consistent with the dual superconducting picture of QCD [64–66]. The Y-ansatz describes the leading string effect and can be successful for parametrization the large distance lattice data of the confining potential [29, 34] at zero temperature.

We summarize the motivation to discuss an effective Y-string model versus the lattice data at high temperature in the following main points: The linear growth property of the confining flux-tube at high temperature has been verified on the lattice [21, 49, 60–63], no substantial changes [67] in the nature of the confining thin-tubes between a quark-antiquark pair on large distance scales, the Y-model seems consistent with lattice data corresponding to the confining potential at $T = 0$ [34]. The expectations that the observed features of the gluonic distribution may arise as a result of the vibration of this underlying Y-shaped string system.

In the Y-baryonic string model, the quarks are connected by three strings that meet at a junction (Fig. 1). The classical configuration corresponds to the minimal area of the string world sheets. Each string's world-sheet (blade) consists of a static quark line and the world-line of the fluctuating junction Fig. 2.

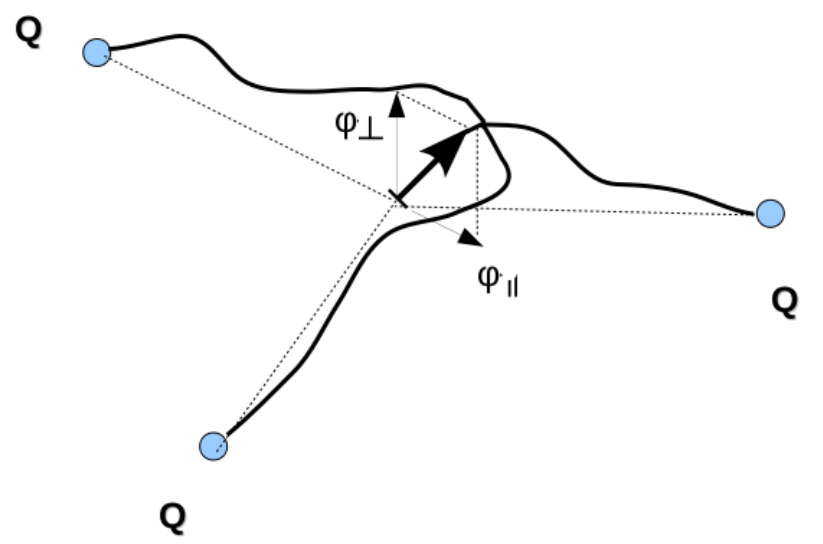

FIG. 1: Fluctuating Y-shaped flux tubes arrangement of three static color sources Q. The junction position is described by the collective coordinate $\varphi$.

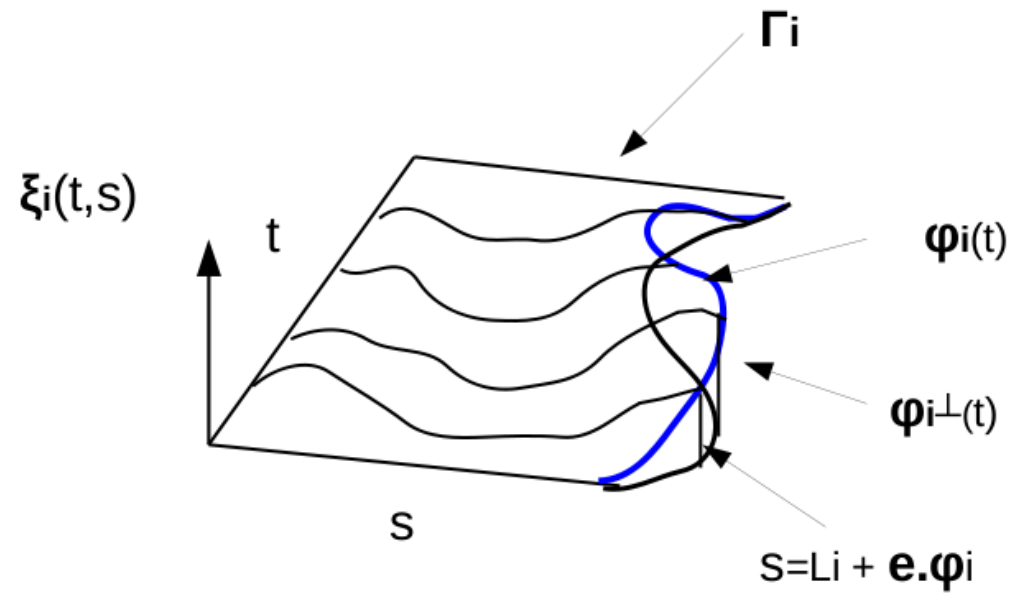

FIG. 2: World sheet traced by one of the strings up to the junction position.

The parameter $s$ and $t$ (time) label the position on string world-sheet (blade) $i$. The position of the junction is given by $s = L_i + e_i.\varphi(t)$. The transverse fluctuations $\boldsymbol{\xi}_i(t,s)$ vanish at the location of the quarks ($s = 0$), and are periodic in the time $t$, with period $1/L_T$(see Fig. 2).

The most natural choice for the string action $S$ is the Nambu–Goto action which is proportional to the surface area

$$S[X] = \sigma \int d\zeta_1 \int d\zeta_2 \sqrt{g}, \tag{4}$$

where $g_{\alpha\beta}$ is the two dimensional induced metric on the blade world sheet embedded in the background $\mathbb{R}^4$,

$$g_{\alpha\beta} = \frac{\partial X}{\partial \zeta_\alpha} \cdot \frac{\partial X}{\partial \zeta_\beta}, \quad (\alpha, \beta = 1, 2),$$
$$g = \det(g_{\alpha\beta}).$$

The vector $X^\mu(\zeta_1, \zeta_2)$ maps the region $\mathcal{C} \subset R^2$ into $R^4$, where gauge fixing is required for the path integrals involving the string partition functions to be well defined with respect to Weyl and re-parametrization invariances. The physical gauge, $X^1 = \zeta_1, X^4 = \zeta_2$ would restrict the string fluctuations to transverse directions to $\mathcal{C}$. In the Quantum level, Weyl invariance is broken in 4 dimensions, however, the anomaly is known to vanish at large distances [7]. The transverse fluctuations $\boldsymbol{X}_\perp = \boldsymbol{\xi}^\mu(t, s)$ vanish at the location of the quarks ($s = 0$), and are periodic in the time $t$, with period $L_T$, that is, Dirchlet boundary condition in addition to the boundary condition from the continuity of the transverse fluctuations $\boldsymbol{\xi}_i(t, s)$

$$\boldsymbol{\xi}_i(t, L_i + \boldsymbol{e}_i \cdot \boldsymbol{\varphi}(t)) = \boldsymbol{\varphi}_{\perp i}(t)\,, \tag{5}$$

The NG action after gauge-fixing and expanding around the equilibrium configuration yields

$$S_{\mathrm{Fluct}} = \sigma L_Y L_T + \frac{\sigma}{2} \sum_{i,j} \int_{\Gamma_i} d^2\zeta \, \frac{\partial \boldsymbol{\xi}_i}{\partial \zeta_j} \cdot \frac{\partial \boldsymbol{\xi}_i}{\partial \zeta_j}, \tag{6}$$

where, $L_Y = \sum_i L_i$ above denotes the total string length. In this model, the junction is assumed to acquire a self-energy term $m$. This results in an additional boundary term to NG action

$$S = S_{\mathrm{Fluct}} + S_{\mathrm{Boundary}},$$

with a static energy and a kinetic energy terms of junction defined as

$$S_{\mathrm{Boundary}} = \left( m L_T + \frac{m}{2} \int_0^{L_T} dt \, |\dot{\boldsymbol{\varphi}}|^2 \right).$$

respectively.

The system's partition function then reads

$$Z = e^{-(\sigma L_Y + m) L_T} \int \mathcal{D}\boldsymbol{\varphi} \, \exp\left( -\frac{m}{2} \int dt \, |\dot{\boldsymbol{\varphi}}|^2 \right) \prod_{i=1}^{3} Z_i(\boldsymbol{\varphi}), . \tag{7}$$

where $Z_i(\boldsymbol{\varphi})$ denotes the partition function for the fluctuations of a given blade that is bounded by the junction worldline $\boldsymbol{\varphi}(t)$:

$$Z_i(\boldsymbol{\varphi}) = \int_{\boldsymbol{\varphi}} \mathcal{D}\boldsymbol{\xi}_i \, \exp\left( -\frac{\sigma}{2} \int |\partial \boldsymbol{\xi}_i|^2 \right). \tag{8}$$

The string partition functions $Z_i(\boldsymbol{\varphi})$ are Gaussian functional integrals and can be calculated according to

$$Z_i(\boldsymbol{\varphi}) = e^{-\frac{\sigma}{2} \int |\partial \boldsymbol{\xi}_{\min,i}|^2} |\det(-\triangle_{\Gamma_i})|^{-(D-2)/2} \tag{9}$$

where $\boldsymbol{\xi}_{\min,i}$ is the minimal-area solution for given $\boldsymbol{\varphi}(t)$. $\triangle_{\Gamma_i}$ denotes the Laplacian acting on the domain (blade) $\Gamma_i$. $\boldsymbol{\xi}_{\min,i}(t, s)$ is harmonic and satisfies the boundary conditions Eq. (5) [29].

Jahn and de Forcrand [29] calculated the baryonic potential, $\mathrm{V}_{qqq}$, by evaluating the determinant of the Laplacian in Eq. (9) based on conformally mapping generalized domains of the blade [29].

The authors in Ref [30] extended the calculations of the above model to the thickness of the fluctuating baryonic junction

$$\langle \boldsymbol{\varphi}^2 \rangle = \frac{\int \mathcal{D}\boldsymbol{\varphi} \, \boldsymbol{\varphi}^2 e^{-S}}{\int \mathcal{D}\boldsymbol{\varphi} \, e^{-S}}\,. \tag{10}$$

The integral over $\boldsymbol{\varphi}$ has been decomposed in Eq. (10) using into parallel and perpendicular components $|\boldsymbol{\varphi}_{w,\perp i}|^2 = |\boldsymbol{\varphi}_w|^2 - |\boldsymbol{\varphi}_w \cdot \boldsymbol{e}_i|^2$ to the plane of the quarks (see Fig ). This has resulted in an expression for the mean-square value of the perpendicular fluctuations [30] given by

$$\langle \boldsymbol{\varphi}_\perp^2 \rangle = \frac{2}{L_T} \sum_{w>0} \frac{1}{m\, w^2 + \sigma w \sum_i \coth(w\, L_i)}, \tag{11}$$

with $w = 2\pi n / L_T$.

The above equation is consistent with the mesonic string fluctuations in the limit $L_T \to \infty$. This can be shown by dividing the string connecting a quark and antiquark into two parts of equal length connected in the middle by a junction, $L = 2L' = 2R$, see Fig. 3. In the limit $L_T \to \infty$, for general $n$ string system of identical lengths $L = L_i$, the perpendicular contribution would then read

$$\langle \boldsymbol{\varphi}_\perp^2 \rangle = \frac{D-n}{\pi n} \int_0^\infty dw \, \frac{1}{m w^2 + n \sigma w \coth(w L)}\,. \tag{12}$$

To a leading order, the integral simplifies to

$$\langle \boldsymbol{\varphi}_\perp^2 \rangle = \frac{D-n}{n\, \pi \sigma} \ln \frac{L}{L_0}\,. \tag{13}$$

The arbitrary constant is contained in $L_0$ which generally will depend on the dimension and the number of strings and the ultraviolet properties of the corresponding gauge model [49]. This indicates that the width of the junction, orthogonal to the plane swept by the quarks, grows logarithmically with the distance.

One can relate Eq. (14) to the mean-square width, $W_0^2$, determined at the symmetry point of the string world sheet [18]

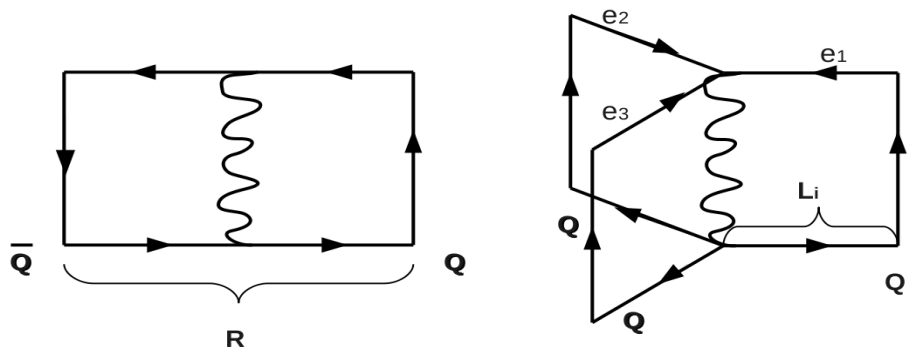

FIG. 3: The world sheets of the strings in a baryon and a meson. The string in the static meson is modeled as being composed of two strings connected by a junction in the middle.

$$W_0^2 = \frac{(D-2)}{2\pi\sigma} \ln \frac{R}{R_0} \,. \tag{14}$$

These equations coincide provided that the constants are identified as $L_0 = 2R_0$. The parameter $m$ which has been absorbed into $L_0$, therefore, scales linearly with the parameter $R_0$.

For our further purpose, the approximation in Eq. (11) can be improved further by including the high temperature effects using a simple convolution with the first order fluctuations $\varphi \rightarrow \int_{-\infty}^{\infty} \phi(\tau)\psi(t-\tau)d\tau$, that is, the fluctuating side of the general domains $\Gamma_a$ describing the world sheet of each blade can be smoothed with a scalar function $\psi$ such that the conformal mapping [29] to a rectangle would read

$$f_i(z) = z + \frac{1}{\sqrt{L_T}} \sum_{\omega=0} \frac{e_k.\varphi_\omega \psi(\omega, L_i)}{\sinh(\omega L_i)}\, e^{\omega z}. \tag{15}$$

The integration over Fourier modes of the fluctuating junction $\phi$ can be performed in a similar way as detailed in Ref. [30]. The mean-square width of the perpendicular fluctuation of the junction acquires a simple modification after solving for the position of the junction $\boldsymbol{\xi}_{min,i}$ for each blade with the convoluted position

$$\boldsymbol{\xi}_{min,i} = \frac{1}{\sqrt{L_T}} \sum_{w} \boldsymbol{\varphi}_{w,\perp i}\, \psi(w, L_i) \frac{\sinh(ws)}{\sinh(wL_i)} e^{iwt} \,. \tag{16}$$

Following the same procedure as Ref. [30] for the calculation of the thickness of the junction (see Appendix. A), the perpendicular fluctuations of Eq. (11) become

$$\langle \varphi_z^2 \rangle = \frac{2}{L_T} \sum_{w>0} \frac{1}{mw^2 + \sigma w \sum_i \coth(wL_i)\psi(w, L_i)}. \tag{17}$$

The form of this convolution scalar can be derived from the mesonic limit. The mesonic limit derived in Ref. [60] would read,

$$w^2(\xi_1, \tau) = \frac{1}{\pi\sigma} \log(\frac{R}{R_0}) + \frac{1}{\pi\sigma} \log\big|\, \chi(\tau)\big|, \tag{18}$$

with $\chi(\tau) = \dfrac{\theta_2(0;\tau)}{\theta_1'(0;\tau)}$. Equating both expressions of Eq. (17) and Eq. (18), with $R = 2L_i = L$, expanding the logarithm in the right hand side and solving for $\psi(w, L_i)$ yields for the smoothing $\psi$ the following expression

$$\psi(w, L_i) = \frac{-kw}{2\sigma \coth(wL_i)} - \frac{(wL_T - \pi)}{2wL_T \coth(wL_i)} \left( \frac{2L_i\, \chi(\tau_i) + 1}{2L_i\, \chi(\tau_i) - 1} \right)^{w\, L_T/\pi - 1} \,. \tag{19}$$

As indicated above, the parameter $m$ shifts the mean-square width of the fluctuations by a constant. The parameter $m$ can be chosen such that $R_0$ cancels out from both sides of Eqs. (17) and (18).

After plugging the smoothing scalar $\psi$ into the mean-square width of the in-plane fluctuations and orthogonalizing the corresponding path integral (Appendix. A), we have

$$\begin{aligned} \langle \varphi_{x,\parallel}^2 \rangle &= \frac{2}{L_T} \sum_{w>0} \frac{1}{A_{x,w} + A_{y,w} - (A_{xy,w}^2 + (A_{x,w} - A_{y,w})^2)^{1/2}}, \\ \langle \varphi_{y,\parallel}^2 \rangle &= \frac{2}{L_T} \sum_{w>0} \frac{1}{A_{x,w} + A_{y,w} + (A_{xy,w}^2 + (A_{x,w} - A_{y,w})^2)^{1/2}}, \end{aligned} \tag{20}$$

where $A_x, A_y$ and $A_{\mathrm{Re}}$ are defined as

$$
\begin{aligned}
A_x &= \left( m w^2 + \sigma w \sum_i \coth(wL_i)\psi(w,L_i) \right) + \left( \frac{\sigma}{2} w + \frac{w^3}{12\pi} \right) \left[ \sum_a e_{a,x}^2 \coth(wL_a)\psi(w,L_i) \right] , \\
A_y &= \left( m w^2 + \sigma w \sum_i \coth(wL_i)\psi(w,L_i) \right) + \left( \frac{\sigma}{2} w + \frac{w^3}{12\pi} \right) \left[ \sum_a e_{a,y}^2 \coth(wL_a)\psi(w,L_i) \right] , \\
A_{\mathrm{xy}} &= \left( \frac{\sigma}{2} w + \frac{w^3}{12\pi} \right) \left[ \sum_a e_{a,x} e_{a,y} \coth(wL_a)\psi(w,L_i) \right] .
\end{aligned}
\tag{21}
$$

## IV. THE GLUONIC PROFILE AND BARYONIC STRINGS

In the next two section we show an analysis of the lattice data from two points of view. In the first subsection, we give a qualitative description of the rendered action density profile. We show how the aspects of the distribution are consistent with the stringlike behavior. We directly compare the width profile of the action density with the string model fluctuations Eq. (17) and Eq. (21) in the following subsection.

### A. Qualitative features

The scheme of measurements has been described in section(II). The lattice operator which characterizes the gluonic field $\mathcal{C}$ is usually taken as the correlation between the vacuum lattice action-density $S(\vec{\rho},t)$ operator, and a gauge-invariant operator representing the heavy baryon state, that is, three Polyakov lines Eq. (2). We take our measurements with a three-loop field-strength tensor according to

$$
F_{\mu\nu}^{\mathrm{Imp}} = \sum_{i=1}^{3} w_i \, C_{\mu\nu}^{(i,i)}, \tag{22}
$$

where $C^{(i,i)}$ is a combination of Wilson loop terms corresponding to loops with lattice extent $i$ used to construct the clover term and $w_i$ are the corresponding weights [68].

Different possible components of the field-strength tensor in Eq. (22) can separately measure the chromo-electric and magnetic components of the flux. The action density, however, is related to the chromo-electromagnetic fields via $\frac{1}{2}(E^2 - B^2)$ and is the quantity of direct relevance to the comparison with the string fluctuations Eqs. (17) and (21).
The reconstructed action density

$$
S(\vec{\rho}) = \beta \sum_{\mu>\nu} \frac{1}{2} \mathrm{Tr}(F_{\mu\nu}^{\mathrm{Imp}})^2 \tag{23}
$$

is accordingly measured on 80 sweeps of stout link smearing. The action-density operator is calculated through an $\mathcal{O}(a^4)$ improved lattice version of the continuum field-strength tensor Eq. (22). The correlation function Eq. (2) is found $\mathcal{C}(\vec{\rho}) = 0$ away from the quark position.

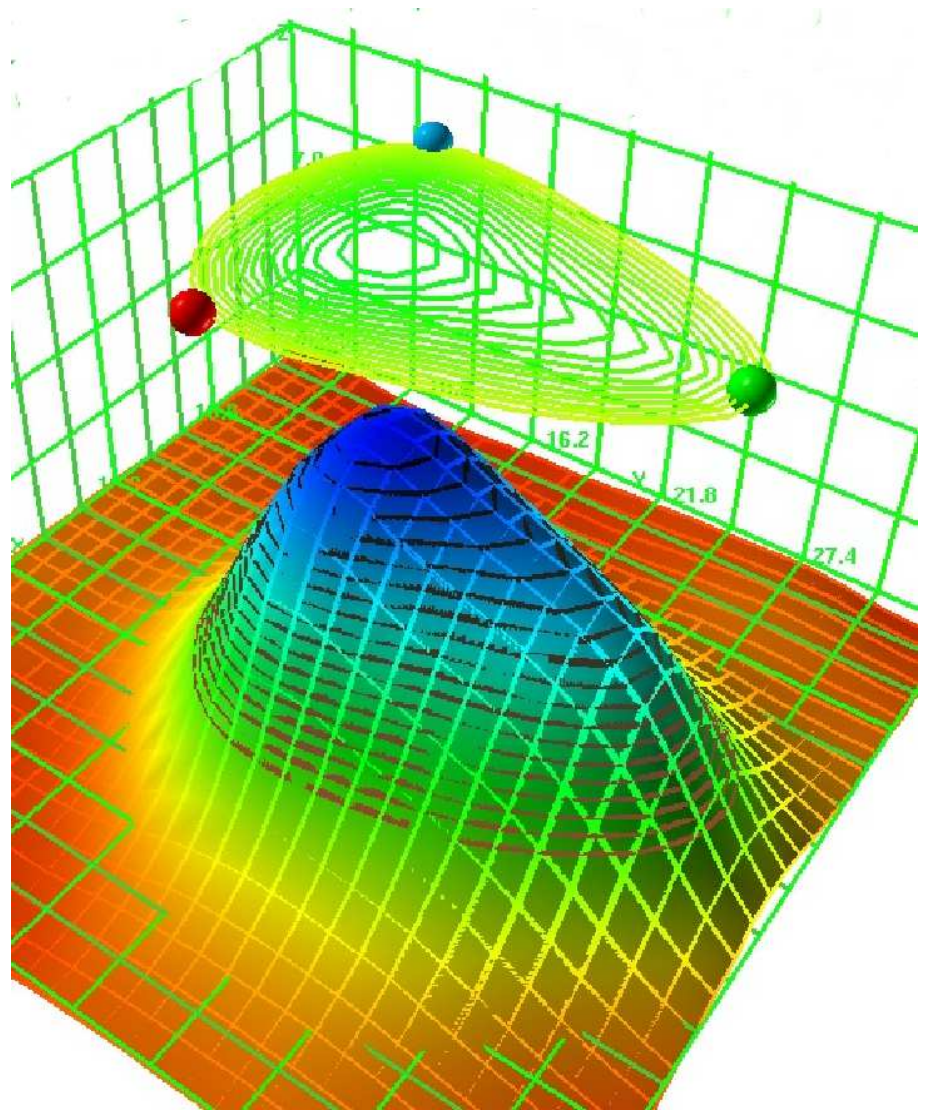

FIG. 4: Surface plot of the flux density surface in the quark plane, $\rho(x,y,z=0)$, together with contour lines. These measurements are taken for an isosceles quark geometry of base A=0.8, height R=1.2 fm and temperature $T = 0.9T_c$.

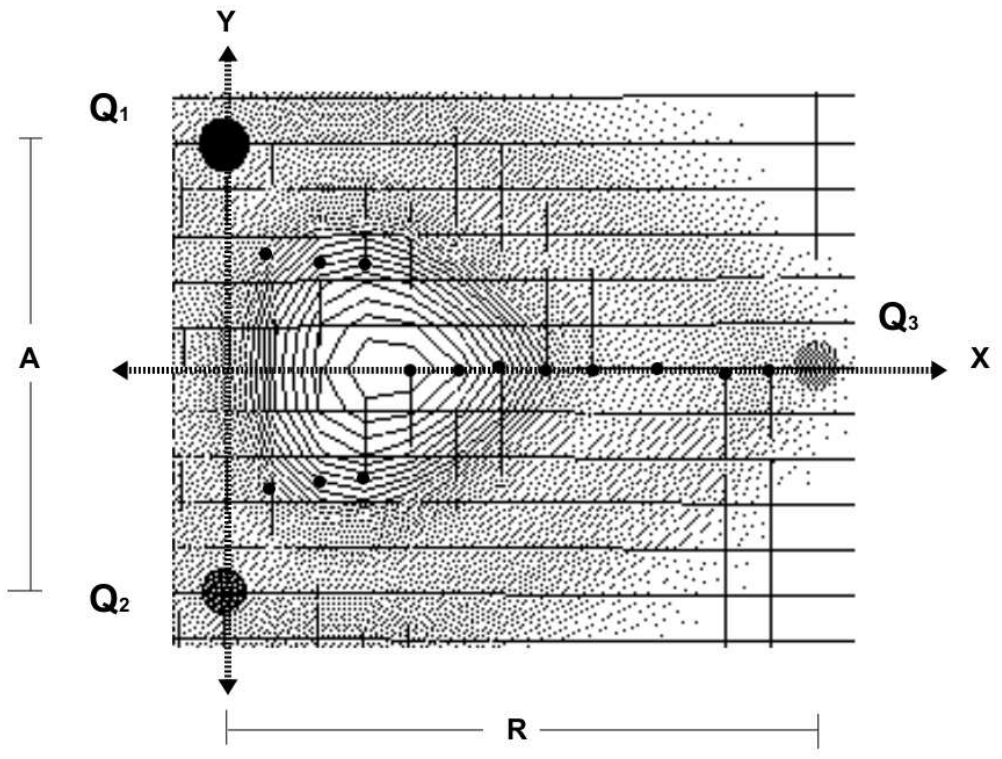

FIG. 5: The contour map of the action density at $T/T_c = 0.8$. The distance between $Q_1$ and $Q_2$ at the base of the triangle is denoted "A", and "R" is the distance between the third quark $Q_3$ to the base. The dots denote the locus of the center of two Gaussians Eq. 24

The surface plot of the scaled flux distribution in the quark plane, $\rho(x, y, z = 0)$, together with contour lines are plotted in Fig. 4. The measurements are taken for an isosceles quark geometry of base A=0.8, height R=1.2 fm and temperature $T = 0.9T_c$. The contour lines are projected onto the surface plot. The density of the contour lines increases near the edges in accord with the gradient of the density scalar field along the x-axis. The flux contours corresponding to the highest values of $C$, however, are the innermost lines inside the triangle.

In general, the action density distribution is non-uniformly distributed. The distribution $\mathcal{C}(\vec{\rho}(x, y, z = 0))$ has an action density maximal curve along the middle line $\vec{\rho}(x, y = 0, z = 0)$ between the two quarks $Q_{1,2}$. With the increase of source $Q_3$ separation, the peak point along the maximal curve $\mathcal{C}(\vec{\rho}(x, y = 0, z = 0))$ shows only subtle movement [27, 28]. At large distances, the topology of these density plots does not indicate an action density pattern resembling the shape of tubes that would form around the perimeter of the three quarks. The distribution displays a peak close to the geometrical center of the triangle at large source separation distance. This is consistent with what we expect from the vibrations of three stringlike flux tubes that meet at a junction. The thin Y-shaped flux tube may delocalize away from its classical configuration and span the whole region throughout the bulk of the triangular 3Q arrangement, tracing out a filled $\Delta$ shape of a nonuniform action density distribution with a maximal in the center.

The nature of the forces that binds the nucleon is usually explored directly in lattice simulations via the fit behavior of a prescribed ansatz to the 3Q potential. The $\Delta$ ansatz amounts for a two body force between the quarks proportional to the perimeter of the 3Q triangle with a string tension half that of the corresponding $q\bar{q}$ system. In the Y-ansatz the string tension is, however, the same as the $q\bar{q}$ system. The force is a three-body force and is proportional to the minimal length of the three strings. At zero temperature, the settled results indicate a $\Delta$-ansatz parametrization for small quark separation distances $R < 0.7$ fm, and the Y-ansatz for $0.7 < R < 1.5$ fm [38]. In perturbation theory [69, 70], the breakdown of the two-particle (Coulombic) interaction picture in the short-range happens at two loop when the first genuine three-body force appears.

It can be a point of subtlety, nevertheless, if the parametrization which provides the best possible fits of the data corresponding to the 3Q system potentials and the profile of the flux tubes are thought to be necessarily the same. Our lattice results for the profile of the flux tube [27, 28] indicates a $\Delta$ shaped flux profile at larger distances. In the following we shall show how the $\Delta$ shape flux configuration consists of three overlapping Y-shaped Gaussian flux-tubes"strings".

The analysis of the fit behavior of the action density profile can directly unravel the relevant string-like configuration amongst certain quark configuration. The best choice of the fitting functions should be based on our experience of what the good fit of the action density profile for a single string would look like. For simplicity, we adopt the approximation where the action density distribution due to a delocalization of a single string can fit to a Gaussian form [19]. This is compatible with the degree of accuracy of our lattice data shown previously in the case of the mesonic flux tubes [49], for example.

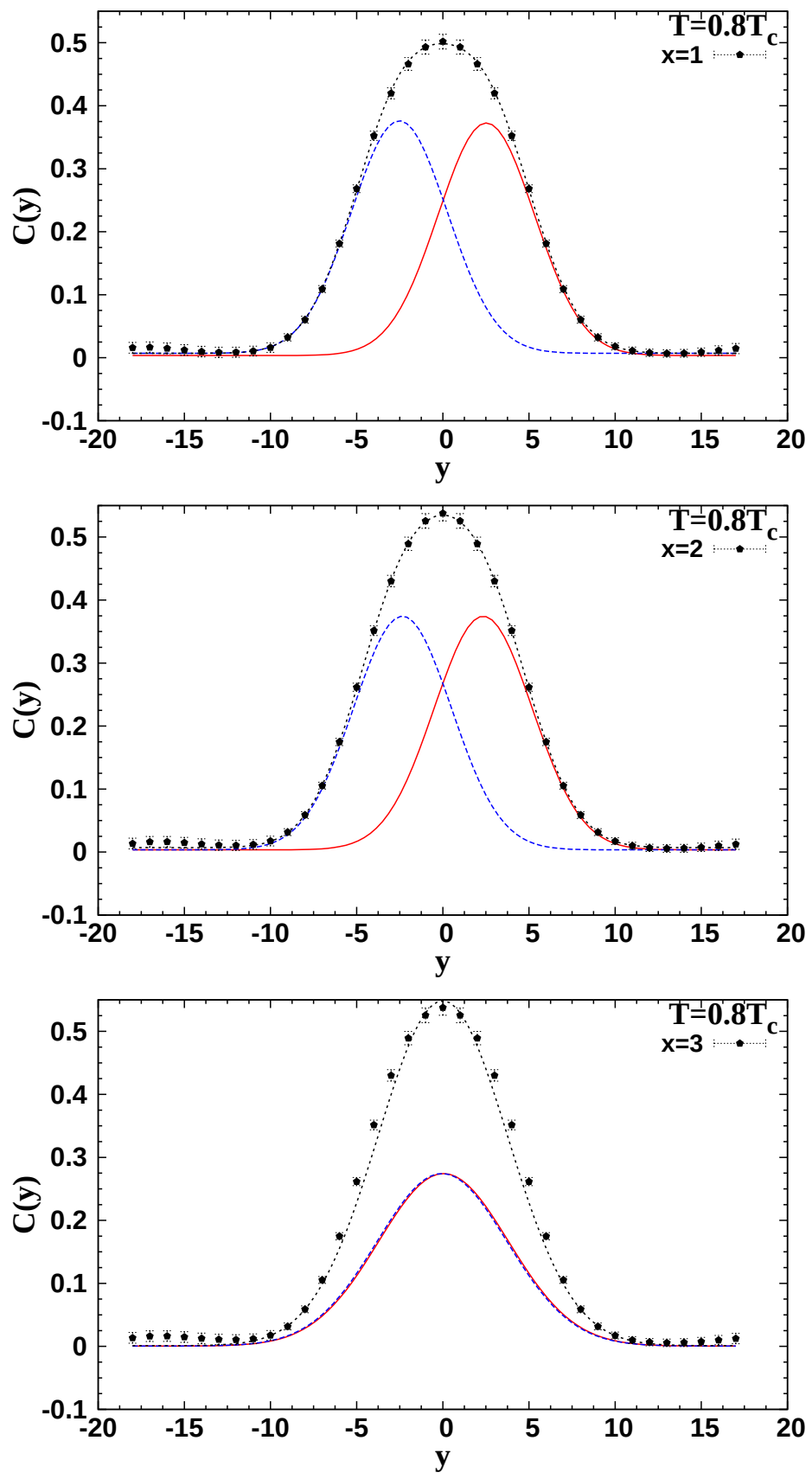

FIG. 6: The density distribution $\mathcal{C}'(\vec{\rho})$ for the isosceles configuration with the base, $A = 1.0$ fm, and height $R = 0.7$ fm at $T/T_c = 0.8$. Data are plotted for the transverse planes $x = 1$ to $x = 3$. The lines correspond to fits of the two Gaussians form Eq. (24) to the density in each plane $\vec{\rho}(x_i, y, 0)$. The distance between the two Gaussians gets closer, and they ultimately coincide at $u(x) = 0$ as we move into farther planes from the base of the triangle.

To unravel the configuration of the strings, we explore the structure of the gluonic distribution with a general ansatz consisting of a two Gaussians

$$G(y; a, w) = A\exp(-(y-u)^2/W^2) + A\exp(-(y+u)^2/W^2). \quad (24)$$

The form assumes a region consisting of a system of two overlapping strings of the same strength $A$, and mean-square width $W^2$. The center of the two strings are separated by distance $|2\,u(x)|$. This form makes use of the symmetry of the flux-tube arrangement and seems suitable to unveil whether the underlying string configuration would be of $\Delta$ or $Y$ form. We scan the gluonic domain with the above fit function for all the distances $x$ from the base $A$ connecting the quarks $Q_1$ and $Q_2$. That is, the Gaussian fits to the action density data are performed for $x_i$ transverse planes between two sources separated by a distance of $R$ from the base up to the third quark $Q_3$ (see Fig. 5).

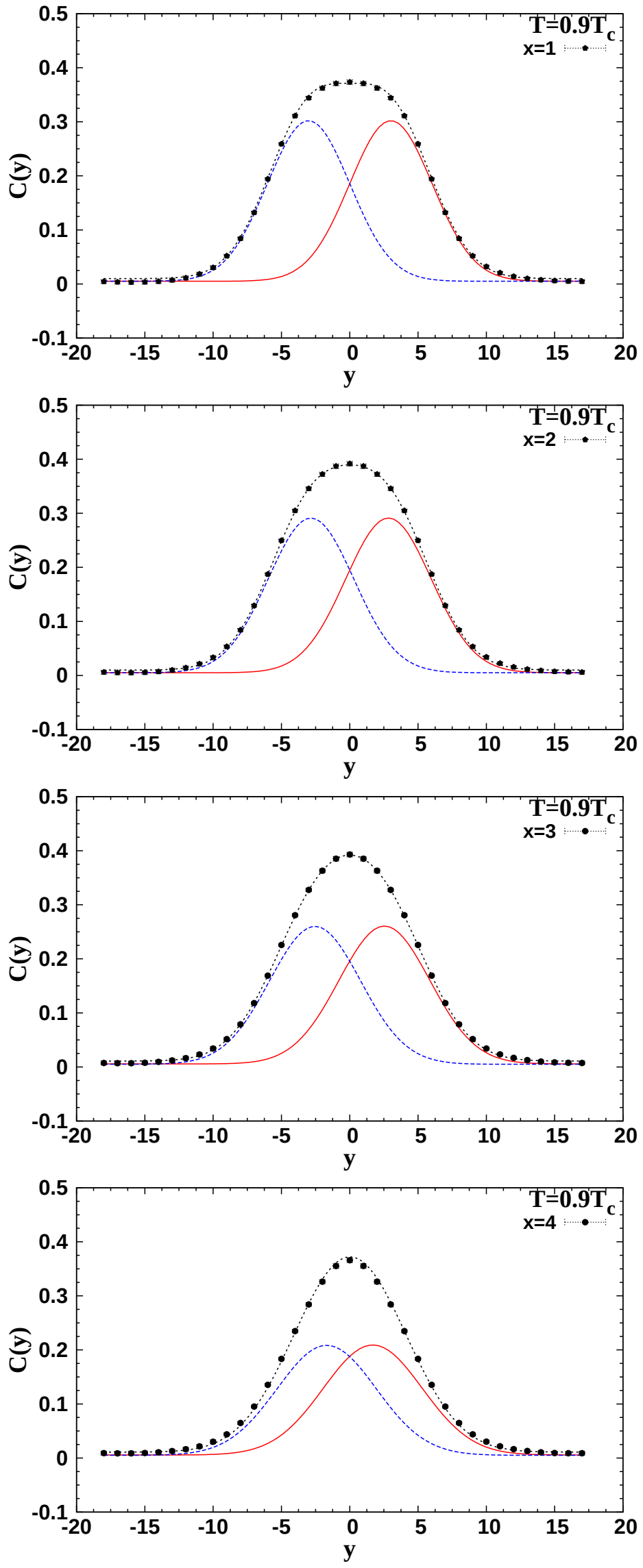

FIG. 7: Same as Fig. 6, the density distribution $\mathcal{C}'(\vec{\rho})$ for the isosceles configuration with the base, $A = 1.0$ fm, and height $R = 0.7$ fm at $T/T_c = 0.9$, respectively. Data are plotted for the transverse planes $x = 1$ to $x = 4$.

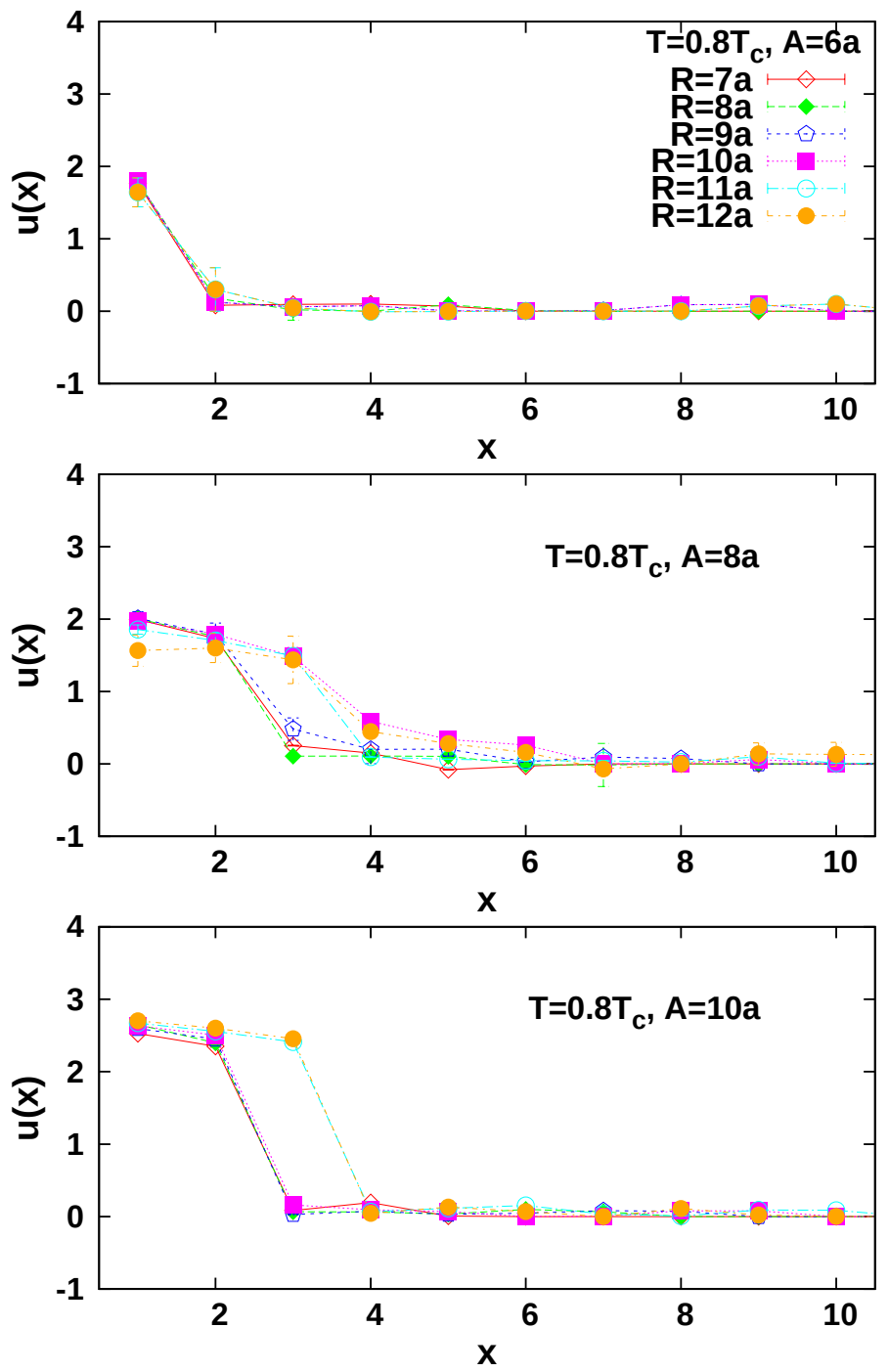

FIG. 8: The separation $u(x)$ between two Gaussians as in Fig. 6 and Fig. 7 used to fit the action density to Eq. (24). Each curve corresponds to $u(x)$ for each third quark $Q_3$ position of the isosceles with the base $A = 6\,a$, $A = 8\,a$ and $A = 10\,a$ at temperature $T/T_c = 0.8$. The legend (in the upper right corner) signifies the third quark position.

The fits of the transverse distribution along the lines $\vec{\rho}(x_i, y, 0)$ are returning good $\chi^2$ to the form Eq. (24). Figures 6 and 7 illustrate the fits to the sum of two Gaussians. The fit parameter $u(x)$ returns non-zero value for fits of the first few planes from the base of the 3Q quark's triangle. The distance between the two Gaussians decreases as we move away towards the third quark position. The interesting behavior of the returned fit parameter $u(x)$ would be at the locus $x_0$ where the separation between the two strings vanishes $u(x_0) = 0$. We refer to this point as the mean-location of the junction which ought to be distinguished from its position in the classical configuration.

The values of the returned fit parameter $u(x_i)$ are plotted in Fig. 8 and Fig. 9 for temperatures $T = 0.8\,T_c$ and $T = 0.9T_c$, respectively. The co-ordinates (Lattice units) are measured from the quark position $x = 0$.

On the other hand, the perpendicular action density $\mathcal{C}'(\vec{\rho}(x_i, 0, z))$ when fitted to Eq. 24 shows no string splitting that is $u(x) = 0$ for all $x \in [0, R]$. Fig. 10 illustrates fits to a single Gaussian form at various planes $x$ in the plane perpendicular to the three quarks at $T/T_c = 0.8$. For the level of accuracy of the data we are analyzing in this investigation, we found that acceptable $\chi^2_{\mathrm{dof}}$ is returned at both temperatures and quark configurations considered here.

At $T/T_c = 0.8$, the two strings show an obvious splitting behavior for the largest isosceles base $A = 10a$ as can be seen in Fig. 8. For this base, the position of the junction $u(x_0) = 0$ interpolates between $x = 3\,a$ (which is the closest plane to the Fermat point of this quark configuration, $x_F = 2.9\,a$) and $x = 4\,a$ for third quark separations $R \geq 1.1$ fm.

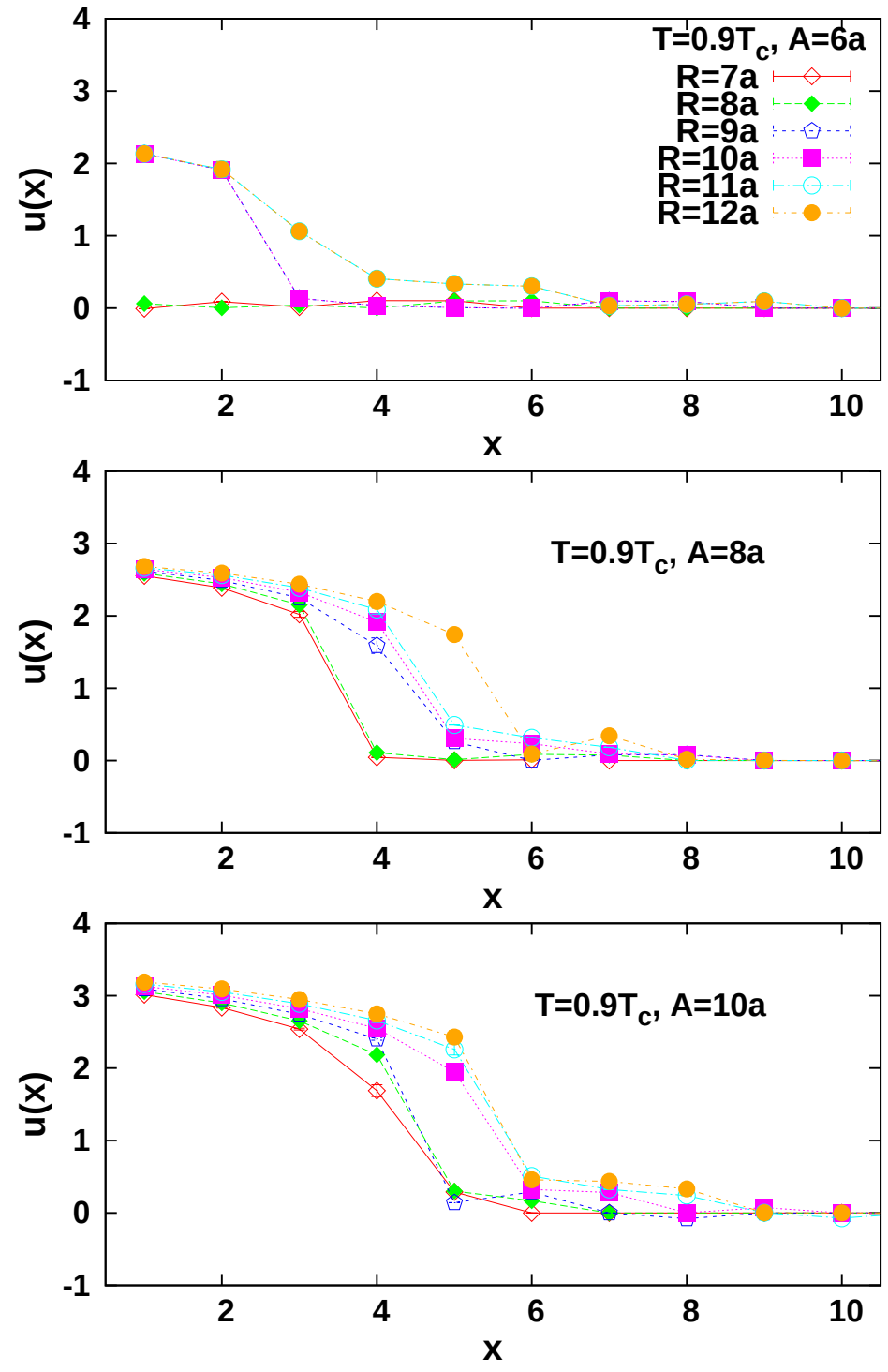

FIG. 9: Similar to Fig. (8) however for temperature $T/T_c = 0.9$. Each curve corresponds to $u(x)$ for each third quark $Q_3$ position of the isosceles with the base $A = 6\,a$, $A = 8\,a$ and $A = 10\,a$. The legend (in the upper right corner) signifies the third quark position.

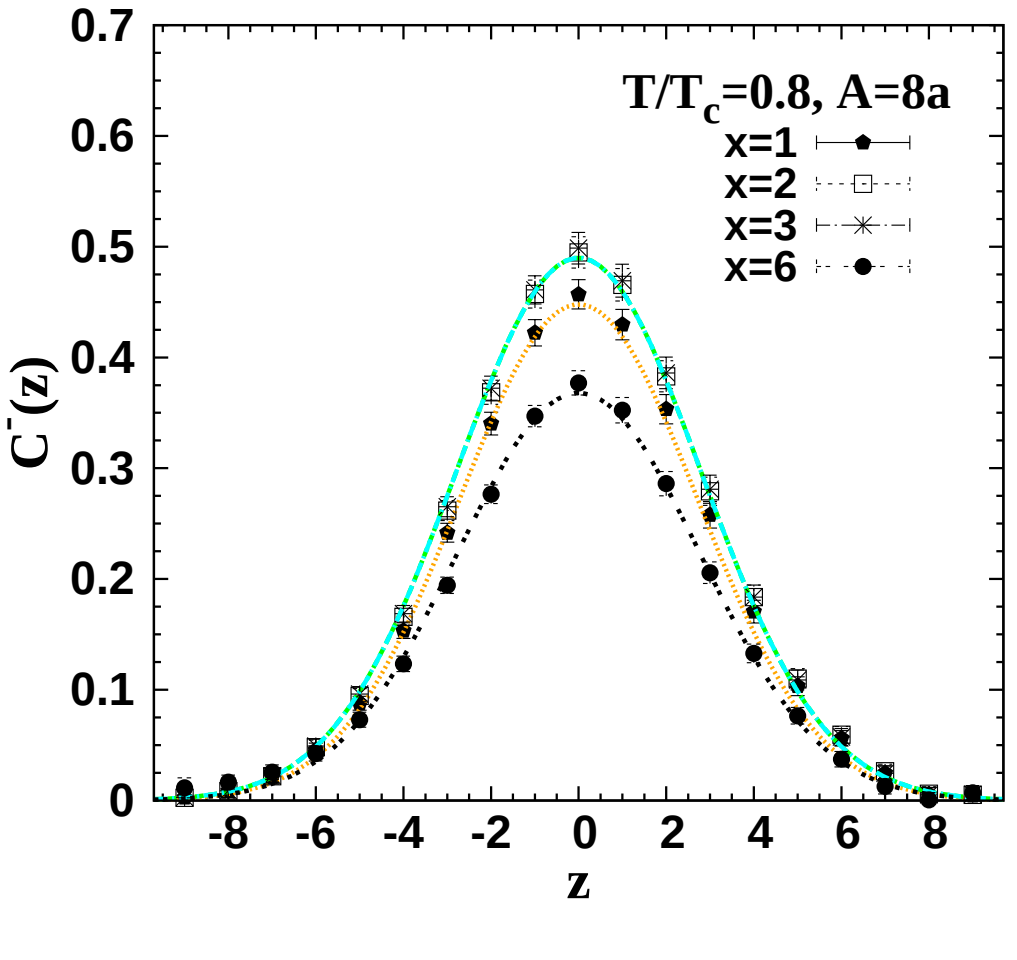

FIG. 10: The density distribution $\mathcal{C}'(\vec{\rho}(x, 0, z))$ in the plane perpendicular to the 3Q plane for an isosceles configuration with the base, $A = 8.0$ fm, and height $R = 0.8$ fm at $T/T_c = 0.8$. The lines show the best fit to a Gaussian form.

The separation between the centers of the two strings manifests for the small bases at the other temperature near the deconfinement point $T = 0.9T_c$. The junction moves towards the third quark as the third quark is pulled away further. At both temperatures, the motion of the mean location $x_0$ of the junction by the increase of the third quark separation, $R \in [3, 12]$, from the base of the isosceles configuration is less or equal 0.2 fm which corresponds roughly to two lattice spacings.

Figure 11 plots the action isolines in the 3Q plane for temperature just before deconfinement point $T = 0.9T_c$. The lines superimposed on the density plot refer to the best fits to the position of the center of each of the two overlapping Gaussians Eq. (24) based on the returned values of the separation parameter $u(x)$, see Figs 6 and 7, for example. At small separation the positions of the junction are very close to the third quark position. The mean position of the strings seem to trace a circumference which looks like a $\Delta$ shape.

At all considered planes, the separation between the two strings, $u(x_i)$, increases with the increase of the temperature, indicating that the profile of the centers of the two Gaussians is consistent with a mean location describing the spread of the gluonic energy with the increase of the temperature and source separation as well.

Figure. 12 compares the profile of the strings for the temperature just before the deconfinement point and the other temperature at the end of the QCD plateau region. For the latter, the junction position is the closer to the Fermat point of the configuration. This indicates that Y-string like system has a maximal length near the deconfinement point.

The second moment, $W_y^2(x)$ is measured by means of the fits to Eq. (24)

$$W_y^2(x_i) = \frac{\int dy\, y^2\, \mathcal{C}'(\vec{\rho}(x_i, y, 0))}{\int dy\, \mathcal{C}'(\vec{\rho}(x_i, y, 0))}. \tag{25}$$

Tables V and VI in Appendix(C) list our measurement of the width of the Flux-tube for the in-plane action density, for the isosceles configurations with base $A = 0.6$ fm, $A = 0.8$ fm and $A = 1.0$ fm at two temperatures $T/T_c = 0.8$ and $T/T_c = 0.9$, respectively. In Ref. [28] we have taken our width measurements using a single Gaussian form. Tables VIII and X show the percentage difference between both width measurements. We note that the difference in width measurements are more pronounced for large quark separations at the higher temperature near the deconfinement point and for the closest planes to the base of the triangle $Q_1$ and $Q_2$.

The transverse profile of the action density fits to a double humped function indicating a system of overlapping strings-like flux tubes. The revealed configurations of these Gaussian flux-tubes show dynamical aspects and reconfigures with respect to the quark configuration and temperature.

Let us point out a third qualitative aspect of the revealed gluonic profile of the 3Q system that can have a stringy character. The gluonic flux in the 3Q system does not exhibit a symmetry between the width measured in the quark plane and that in the perpendicular direction. The flux strength distribution revealed with the action density using the Wilson loop does not appear to produce an asymmetric gluonic pattern. In Ref. [26] the radius of the tube is calculated with cylindrical coordinates assuming a cylindrical symmetry of the tube.

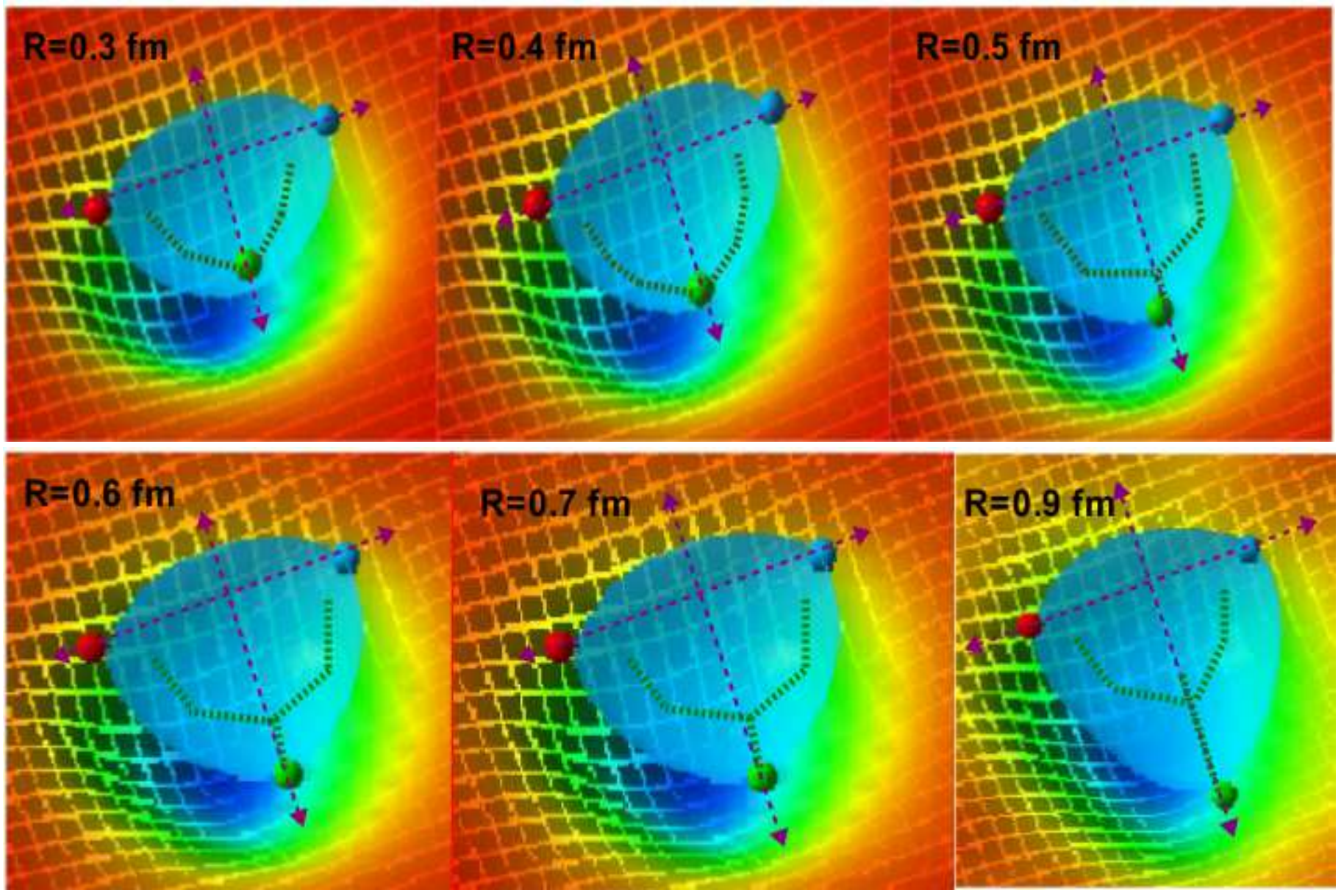

FIG. 11: The action density isolines in the (3Q) plane together with the lines connecting the center position of each Gaussian in the fit Eq. (24). Each sphere denotes the quark position.

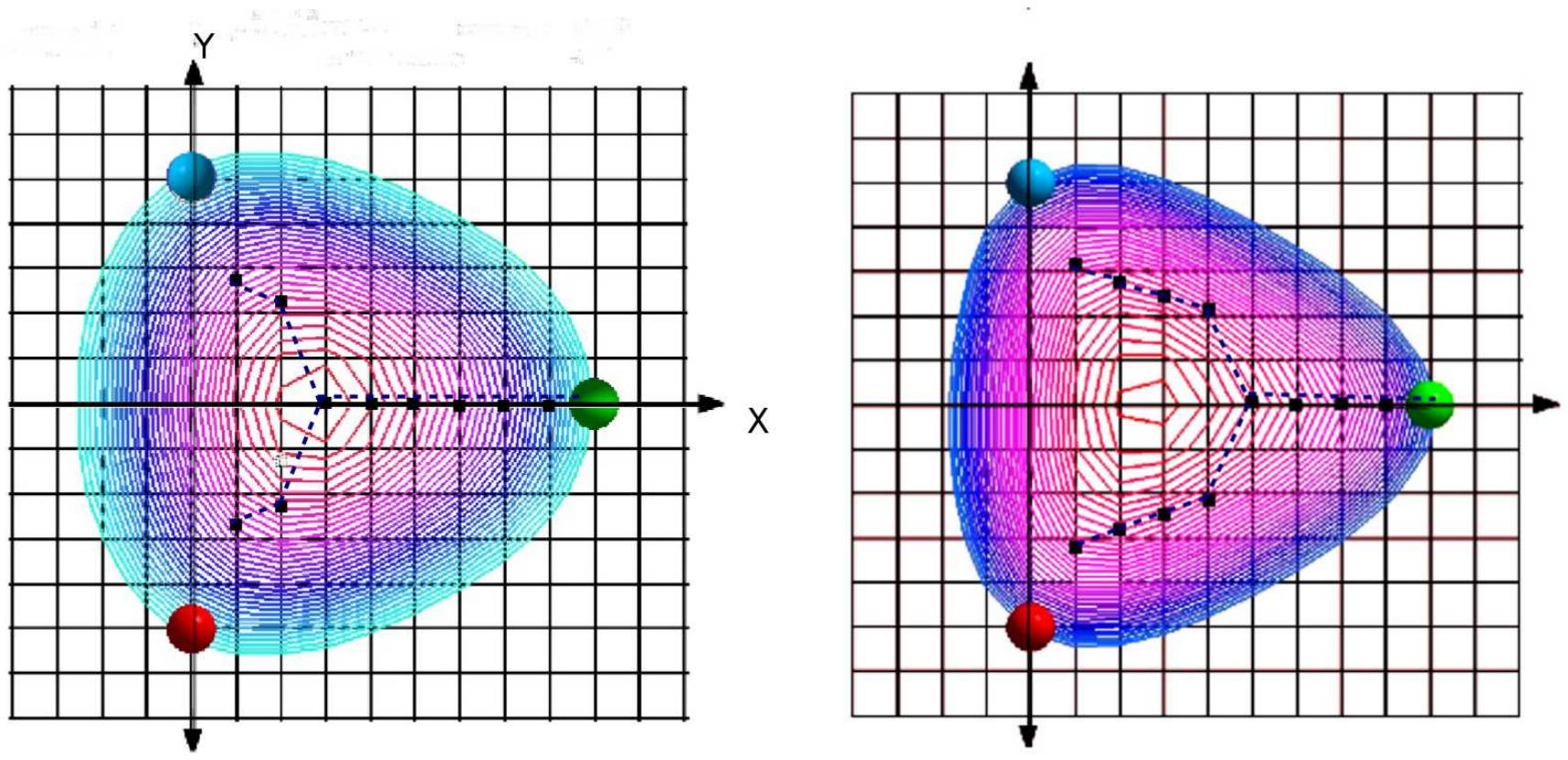

FIG. 12: This plot shows that the notable change on the profile of the baryonic flux arrangement, with the temperature increase, is the movement of the junction to the inner region of the quark configuration.

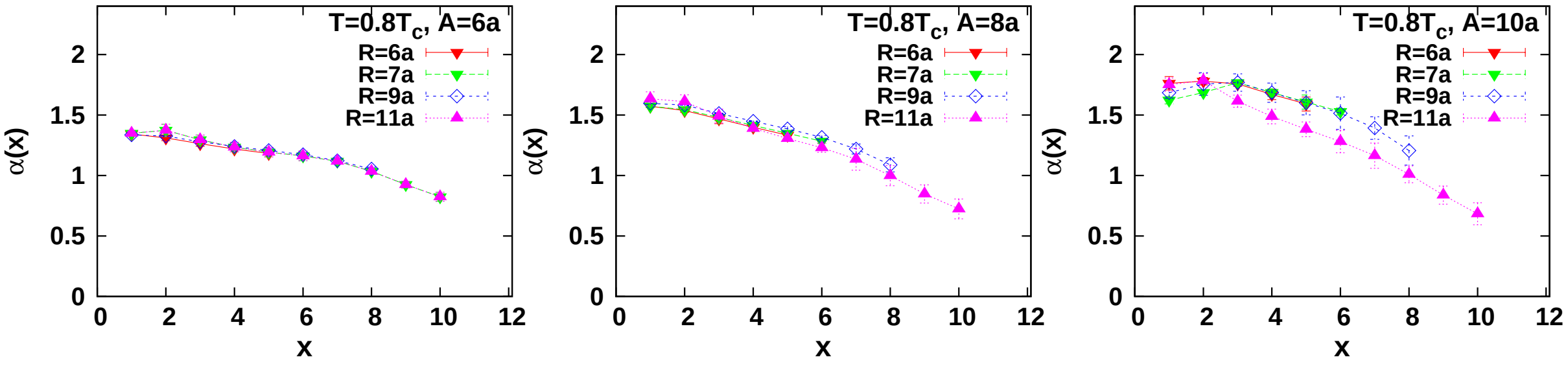

FIG. 13: Comparison of the ratio, $\alpha$, of the mean squared width of the flux parallel and perpendicular to the quarks' plane for three isosceles bases $A = 0.6$ fm, $A = 0.8$ fm and $A = 1.0$ fm at $T/T_c = 0.9$.

The string picture indicates an asymmetry in the mean square width between the two planes [30]. We compute this ratio of the action density in the two perpendicular planes based on fits to the form Eq. (24). The width of the tube in the perpendicular direction is measured through Gaussian fits as

$$W_z^2(x) = \frac{\int d\,z\,z^2\,\mathcal{C}'(\vec{\rho}(x_i,0,z))}{\int d\,z\,\mathcal{C}'(\vec{\rho}(x_i,0,z))}. \qquad (26)$$

Tables VII and IX included in Appendix(C) list our measurement of the width of the Flux-tube for the perpendicular plane action density, for the isosceles configurations with base $A = 0.6$ fm, $A = 0.8$ fm and $A = 1.0$ fm at two temperatures $T/T_c = 0.8$ and $T/T_c = 0.9$, respectively.

We measure the aspect ratio between the width in the quark plane and that in the perpendicular plane to the quarks according to

$$\alpha(x) \equiv \frac{W_y^2(x)}{W_z^2(x)}. \qquad (27)$$

We plot in Fig. 13 the aspect ratio at the temperature $T/T_c = 0.8$ for the indicated quark configurations. Generally, the aspect ratio is greater than 1 for planes $x \in [0, 8[$ indicating that in-plane fluctuations are greater than the perpendicular fluctuations. This results are consistent with a greater restoring forces in the quark planes for the Y-string-like gluonic distribution.

The aspect ratio increases with the base length and decreases, however, as we move up through the planes to the third quark position. With the increase of the base length parameter of the 3Q triangle, the split between the two strings is larger, giving rise to larger width of the fluctuations. For the perpendicular direction, on the other hand, values in Tables IX show smaller growth in the mean square width with the base compared to the parallel width. The aspect ratio tends to assume closer values to 1 at those planes $x$ larger than 8 lattice units from the base. This is a natural consequence of the reduction in the effects of the junction's system as we move the third quark source farther from the base.

### B. Broadening aspects

In the following the broadening of the flux tube is compared to the corresponding string model predictions. Formulas of Eq. (17) and Eq. (21) shown in the last section account for the tube's mean-square width for both the in-plane and perpendicular directions to the 3Q plane, respectively. A study of the fit behavior of each separate component can provide an indication on the compatibility of the baryonic string model with the measured LGT junction profile.

The Y-string's configuration has to be fixed before proceeding to fits with the lattice data. Figure 14 demonstrates the proposed string configuration with respect to the quark positions. We focus our analysis on the flux tube's action density due to 3Q planar configuration corresponding to isosceles triangles with bases of length $A = 6\,a$, $A = 8\,a$ and $A = 10a$.

The locus of the junction $x_f$ is fixed at the Fermat point of the isosceles triangle, i.e. a point such that the total distance from the three vertices of the triangle to this point is the minimum possible. The isosceles triangular configurations have the property of having the same locus of the Fermat point, the position of the Fermat point does not depend on the height $R$ of the triangle, and is given by $R = A/(2\sqrt{3})$. This planar quark setup is convenient to simplify the study of a baryonic junction on a lattice structure.

The measured values in Table V for the in-plane width profile are indicating a growth in the tube's mean-square width at the first four transverse planes $x = 1$ to $x = 4$ as the third color source $Q_3$ is pulled apart. The growth in the flux-tube width could be compared to the corresponding growth in the junction fluctuation Eq. (17) and also Eq. (21) for the perpendicular fluctuations. Since the junction's fluctuations are non-local, this comparison can be performed by fitting the formula of Eq. 17 and Eq. 21 at each selected transverse plane to the tube's measured widths. Here, we focus our analysis on the first four planes from the base of the quark triangle.

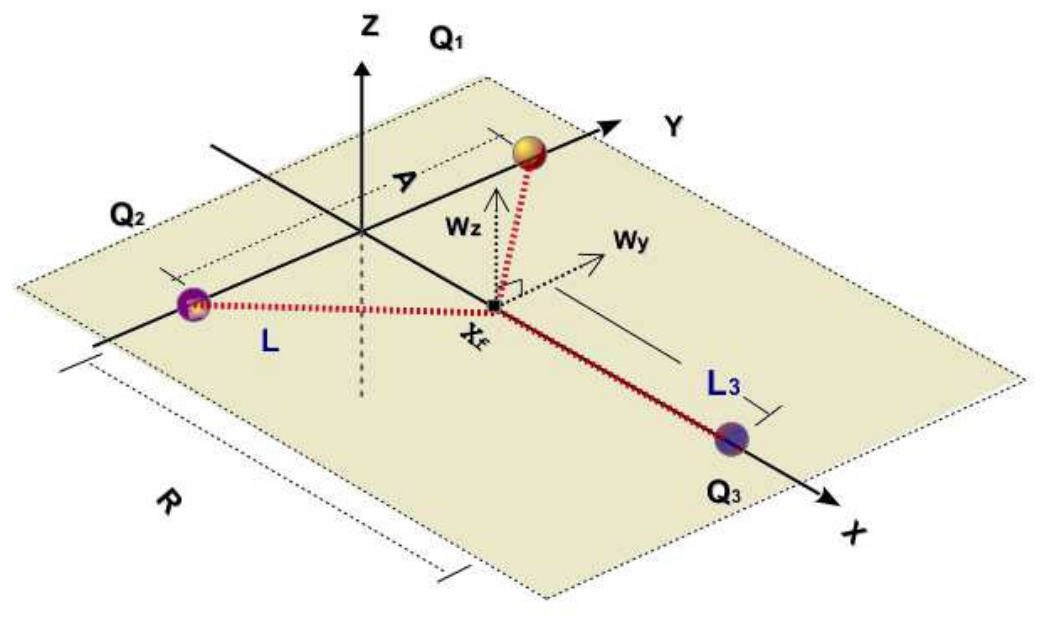

FIG. 14: Schematic diagram shows the position of the quarks and the configuration of the Y-string, the junction's position, $x_f$, is fixed at Fermat point [29, 30].

Let us first fit the measured lattice data for the in-plane width profile $W_x^2$ at the temperature $T/T_c = 0.8$ to Eq. (21). Table I summarizes the returned values of

TABLE I: The returned values of the $\chi^2_{\rm dof}(x)$ corresponding to fits of the in-plane width $W_y^2(x)$ of the action density at each plane $x$ to the string model formula Eq. (21), the fits are for isosceles triangle quark configuration of base $A = 6\,a$, $A = 8\,a$ and $A = 10\,a$ at $T = 0.8\,T_c$.

| (a) $A = 6a$ | | | | | | | |
|---|---|---|---|---|---|---|---|
| Fit range | 4-10 | 4-13 | 5-13 | 6-13 | 7-13 | 8-13 | 9-13 |
| $\chi^2_{\rm dof}(1)$ | 31.8 | 21.5 | 4.9 | 1.38 | 1.03 | 1.2 | 1.1 |
| $\chi^2_{\rm dof}(2)$ | 5.6 | 5.2 | 3.9 | 4.9 | 5.4 | 5.6 | 5.2 |
| $\chi^2_{\rm dof}(3)$ | 64.1 | 44.9 | 7.8 | 4.3 | 5.1 | 5.1 | 4.5 |
| $\chi^2_{\rm dof}(4)$ | | | 3.8 | 3.9 | 4.1 | 4.0 | 3.5 |

| (b) $A = 8a$ | | | | | | | |
|---|---|---|---|---|---|---|---|
| Fit range | 5-9 | 06-10 | 5-12 | 6-12 | 7-12 | 8-12 | 9-12 |
| $\chi^2_{\rm dof}(1)$ | 5.7 | 5.7 | 12.11 | 4.5 | 1.3 | 1.15 | 0.3 |
| $\chi^2_{\rm dof}(2)$ | 25.0 | 25.0 | 64.4 | 19.9 | 5.3 | 1.1 | 0.2 |
| $\chi^2_{\rm dof}(3)$ | 18.1 | 18.1 | 49.2 | 15.5 | 4.9 | 3.2 | 3.4 |
| $\chi^2_{\rm dof}(4)$ | 10.6 | 10.6 | 35.5 | 10.6 | 7.3 | 4.1 | 2.5 |

| (c) $A = 10a$ | | | | | | | |
|---|---|---|---|---|---|---|---|
| Fit range | 05-09 | 5-11 | 6-12 | 7-12 | 8-12 | 9-12 | 10-12 |
| $\chi^2_{\rm dof}(1)$ | 30 | 41.8 | 22.3 | 8.5 | 2.4 | 0.7 | 0.3 |
| $\chi^2_{\rm dof}(2)$ | 47.9 | 21.9 | 15.7 | 1.6 | 0.7 | 0.11 | 0.07 |
| $\chi^2_{\rm dof}(3)$ | 600 | 487 | 152 | 37 | 11 | 2.5 | 0.2 |
| $\chi^2_{\rm dof}(4)$ | 159.9 | 128.9 | 48.3 | 22.1 | 11.05 | 3.7 | 2.1 |

TABLE II: Same as Table I; however the values of the $\chi^2_{\rm dof}$ are returned from the fits of formula Eq. (17) to the perpendicular width of the action density $W_z^2$.

| (a) $A = 6a$ | | | | | | |
|---|---|---|---|---|---|---|
| Fit range | 5-12 | 6-12 | 7-12 | 8-12 | 9-12 | 10-12 |
| $\chi^2_{\rm dof}(1)$ | 1.6 | 1.8 | 1.7 | 1.5 | 1.3 | 0.5 |
| $\chi^2_{\rm dof}(2)$ | 3.2 | 3.3 | 3.1 | 2.6 | 2.1 | 0.8 |
| $\chi^2_{\rm dof}(3)$ | 6.2 | 6.4 | 5.8 | 5.0 | 4.0 | 1.5 |
| $\chi^2_{\rm dof}(4)$ | 10.5 | 10.7 | 9.7 | 8.4 | 6.8 | 2,7 |

| (b) $A = 8a$ | | | | | | |
|---|---|---|---|---|---|---|
| Fit range | 05-12 | 6-12 | 7-12 | 8-12 | 9-12 | 10-12 |
| $\chi^2_{\rm dof}(1)$ | 9.7 | 1.6 | 0.7 | 0.9 | 1.1 | 0.8 |
| $\chi^2_{\rm dof}(2)$ | 5.0 | 1.5 | 1.4 | 1.7 | 1.8 | 1.43 |
| $\chi^2_{\rm dof}(3)$ | 9.2 | 5.6 | 6.6 | 9.7 | 8.5 | 6.1 |
| $\chi^2_{\rm dof}(4)$ | 11.6 | 10.1 | 12.1 | 13.6 | 14.0 | 9.0 |

| (c) $A = 10a$ | | | | | | |
|---|---|---|---|---|---|---|
| Fit range | 5-12 | 6-12 | 7-12 | 8-12 | 9-12 | 10-12 |
| $\chi^2_{\rm dof}(1)$ | 85.1 | 22 | 3.6 | 1.6 | 1.0 | 0.4 |
| $\chi^2_{\rm dof}(2)$ | 77.2 | 15.4 | 2.3 | 0.3 | 0.3 | 0.4 |
| $\chi^2_{\rm dof}(3)$ | 74.2 | 16.0 | 3.0 | 1.0 | 1.4 | 1.4 |
| $\chi^2_{\rm dof}(4)$ | 74.2 | 16.5 | 3.7 | 1.1 | 1.4 | 1.6 |

$\chi^2_{\rm dof}(x)$ from resultant fits to the indicated separation range $R$ at four consecutive transverse planes $x = 1$ to $x = 4$. In general, the fits show strong dependency on the fit range, especially with the inclusion of the points at small $Q_3$ sources separations. Also high values of $\chi^2_{\rm dof}$ is returned when fits include the entire considered range of sources' separations, i.e.,$R = 5a$ to$R = 12a$. However, the values of $\chi^2_{\rm dof}$ rapidly decrease with excluding those points at short distance separations. The fit reaches acceptable values for width measured at the plane $x = 1$ for isosceles triangles' quark configuration with bases of length $A = 6\,a$ at third source $Q_3$ separations $R > 6$ fm. However, the triangular configuration of base lengths $A = 8\,a$ and $A = 10\,a$ returns good values of $\chi^2_{\rm dof}$ for both the plane $x = 1$and $x = 2$. The best fits for the triangle configuration with base lengths $A = 8\,a$ are obtained for $Q_3$ separations $R > 6$ at the plane $x_1$. It is apparent that best fits shift one lattice spacing, that is $x = 2$, when the length of triangle base parameter is increased to $A = 10\,a$, and the best fits is returned with $Q_3$ source separation $R > 7$.

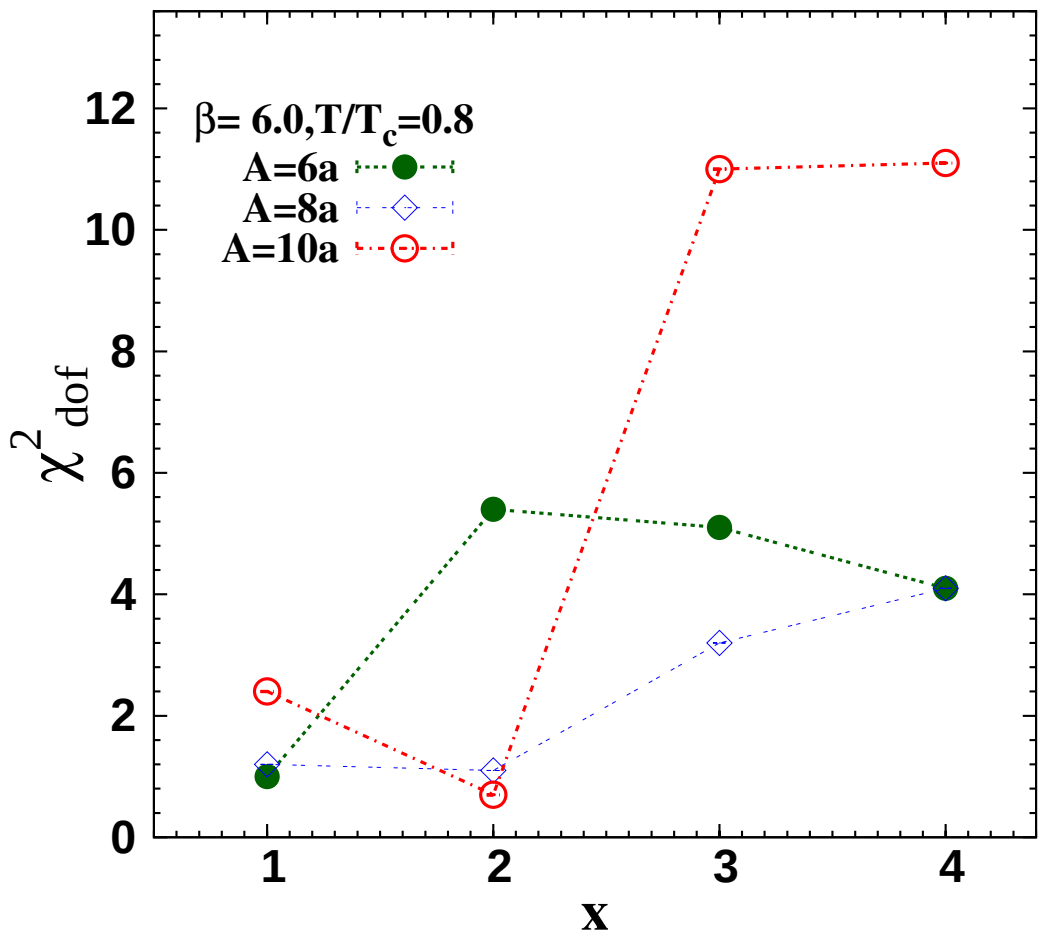

FIG. 15: Plots $\chi^2_{\rm dof}$ of Table I for the isosceles base $A = 0.6\,a$, $A = 0.8\,a$ and $A = 1.0\,a$. The values for the same base length parameter are joined with dashed lines for illustration. The plotted $\chi^2_{\rm dof}$ values are plane $x = 1$ corresponding to the base parameters $A = 0.6\,a$, and $x = 2$ for $A = 8a$ and $A = 10a$.

TABLE III: The returned values of the $\chi^2_{\rm dof}(x)$ corresponding to fits of the in-plane width $W_y^2(x)$ of the action density at each plane $x$ to the string model formula Eq. (17), the fits are for isosceles triangle quark configuration of base $A = 6\,a$, $A = 8\,a$ and $A = 10\,a$ at $T = 0.9\,T_c$.

| (a) $A = 6a$ | | | | | | | |
|---|---|---|---|---|---|---|---|
| Fit range | 05-09 | 05-10 | 05-11 | 05-12 | 06-12 | 07-12 | 08-12 |
| $\chi^2_{\rm dof}(1)$ | 31.76 | 43.24 | 46.2 | 7.79 | 7.97 | 3.42 | 0.42 |
| $\chi^2_{\rm dof}(2)$ | 1.37 | 1.29 | 1.12 | 0.96 | 1.12 | 0.17 | 0.12 |
| $\chi^2_{\rm dof}(3)$ | 10.6 | 9.72 | 8.6 | 8.02 | 5.06 | 4.4 | 1.02 |
| $\chi^2_{\rm dof}(4)$ | 50 | 43.5 | 38.16 | 34.4 | 36.3 | 7.8 | 2.6 |

| (b) $A = 8a$ | | | | | | | |
|---|---|---|---|---|---|---|---|
| Fit range | 05-09 | 05-10 | 07-11 | 05-12 | 06-12 | 07-12 | 08-12 |
| $\chi^2_{\rm dof}(1)$ | 49.9 | 26.5 | 3.4 | 67.8 | 19.8 | 4.0 | 1.4 |
| $\chi^2_{\rm dof}(2)$ | 18.6 | 23.3 | 1.7 | 56.2 | 7.8 | 1.8 | 0.4 |
| $\chi^2_{\rm dof}(3)$ | 5.13 | 22.6 | 1.0 | 52.3 | 21 | 1.2 | 0.6 |
| $\chi^2_{\rm dof}(4)$ | 89 | 111 | 25.2 | 128 | 60 | 30 | 16 |

| (c) $A = 10a$ | | | | | | | |
|---|---|---|---|---|---|---|---|
| Fit range | 05-09 | 07-09 | 07-11 | 06-12 | 07-12 | 08-12 | 09-12 |
| $\chi^2_{\rm dof}(1)$ | 167 | 34.7 | 40.3 | 94.8 | 51 | 12.67 | 6.18 |
| $\chi^2_{\rm dof}(2)$ | 286.2 | 38 | 31.5 | 74.7 | 29 | 11 | 3.96 |
| $\chi^2_{\rm dof}(3)$ | 116.5 | 80 | 51.9 | 73 | 43.2 | 9.6 | 1.02 |
| $\chi^2_{\rm dof}(4)$ | 63.5 | 7.7 | 5.4 | 38.5 | 3.5 | 1.3 | 0.2 |
| $\chi^2_{\rm dof}(5)$ | | 10.5 | 7.2 | 5.6 | 6.3 | 3.1 | 0.9 |

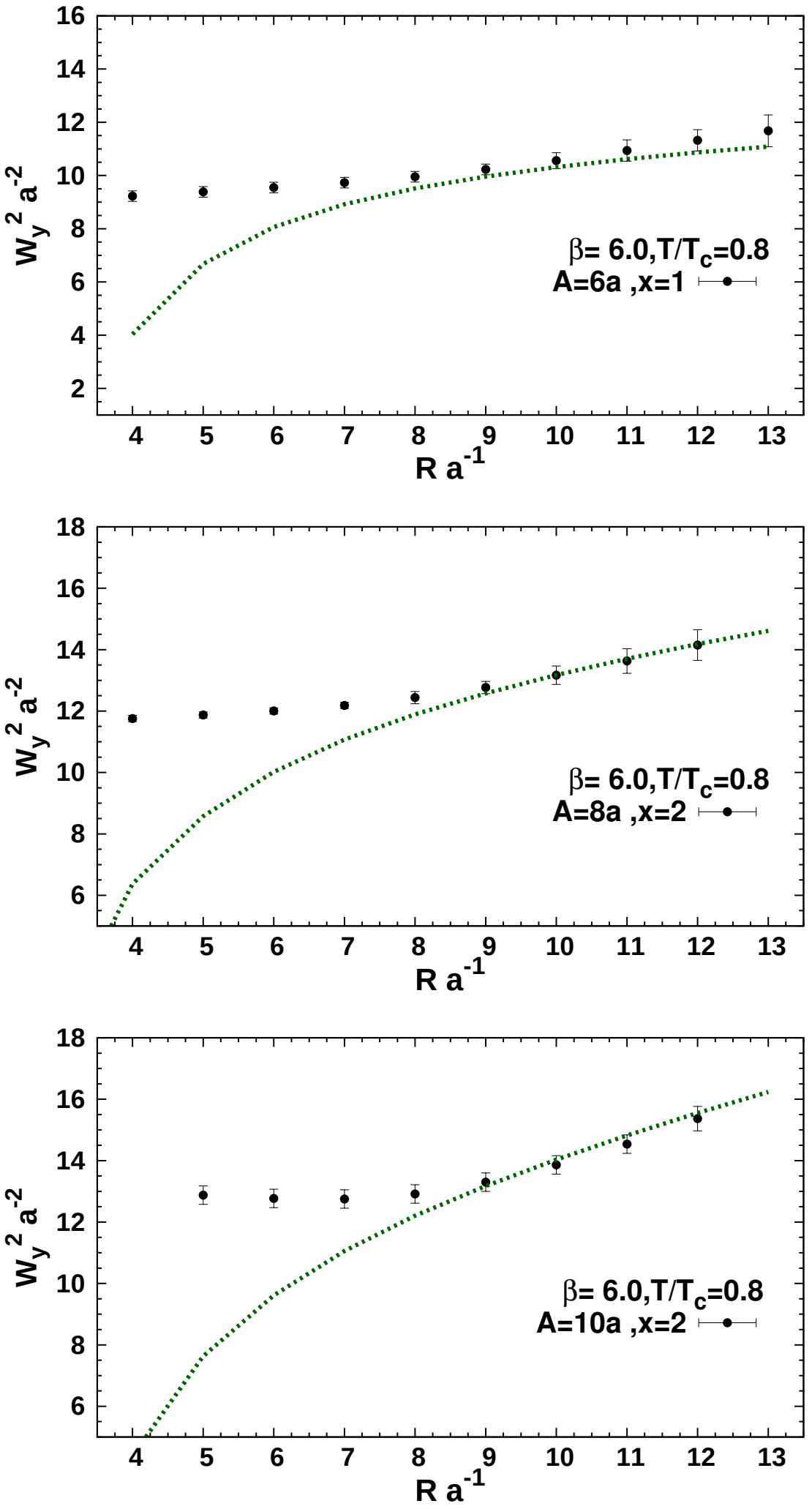

FIG. 16: The in-plane width $W_y^2$ of the action density at temperature $T/T_c = 0.8$ for the depicted transverse plane and the indicated isosceles base $A$. The lines correspond to the width according to the best fits to the string picture formula of Eq. (21)

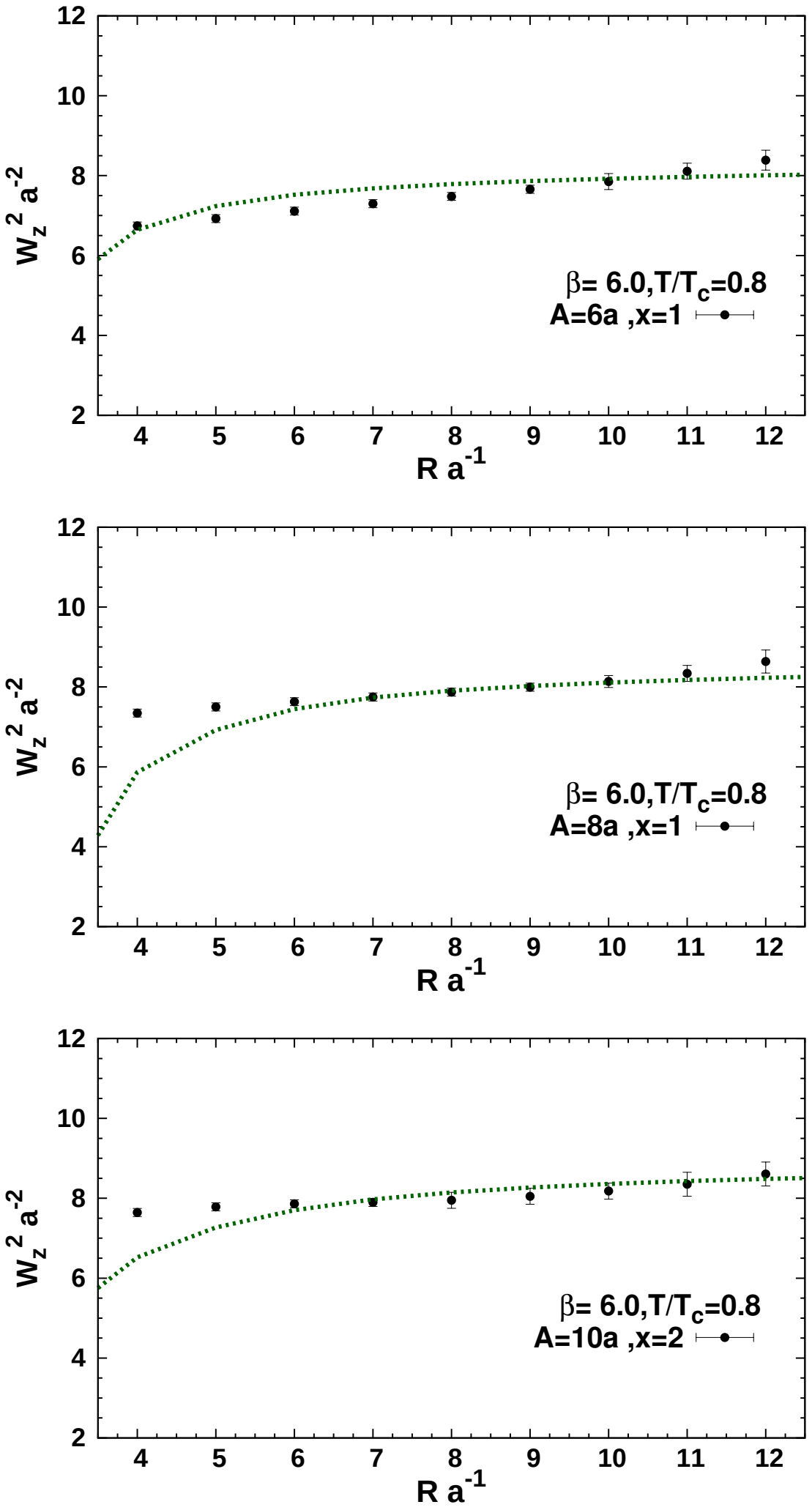

FIG. 17: The perpendicular width $W_z^2$ at temperature $T/T_c = 0.8$ for the depicted transverse planes, $x$, and isosceles base of the corresponding triangular quark arrangement. The lines correspond to the best fits of the string model formula Eq. (17).

Nevertheless, the existence of particular planes at which the above indicated best matches with Formulas of Eq. (21) suggests that some planes may receive a larger contribution of the junction's fluctuations than others. Figure 15 plots $\chi^2_{\rm dof}$ for a selected fit region for planes $x = 1$ to $x = 4$. As mentioned above, the plane at which we obtain the minimal in $\chi^2_{\rm dof}$ depends on the length of base of the triangular isosceles quark configuration.

The positions of the Fermat point of the three isosceles of bases $A = 6a$, $A = 8a$ and $A = 10a$ would be $x_f = 1.7a$, $x_f = 2.3a$ and $x_f = 2.9a$, respectively. Recalling that the Y-string configuration which we fit to lattice data is such that the position of the junction is fixed at the Fermat point at both considered temperatures. One may expect, accordingly, that those planes closer in position to the string's junction (See Fig. 14, the configuration to which we fit the width profile) can provide better fits to the lattice data. We observe, however, that the planes of best fits manifest in accord with the profile of the two Gaussians shown in Figs 8 and 9 rather than junction's classical position at the Fermat point. For the in-plane fluctuations of Eq. (21), the greatest contribution of the junction appears to be in one lattice spacing

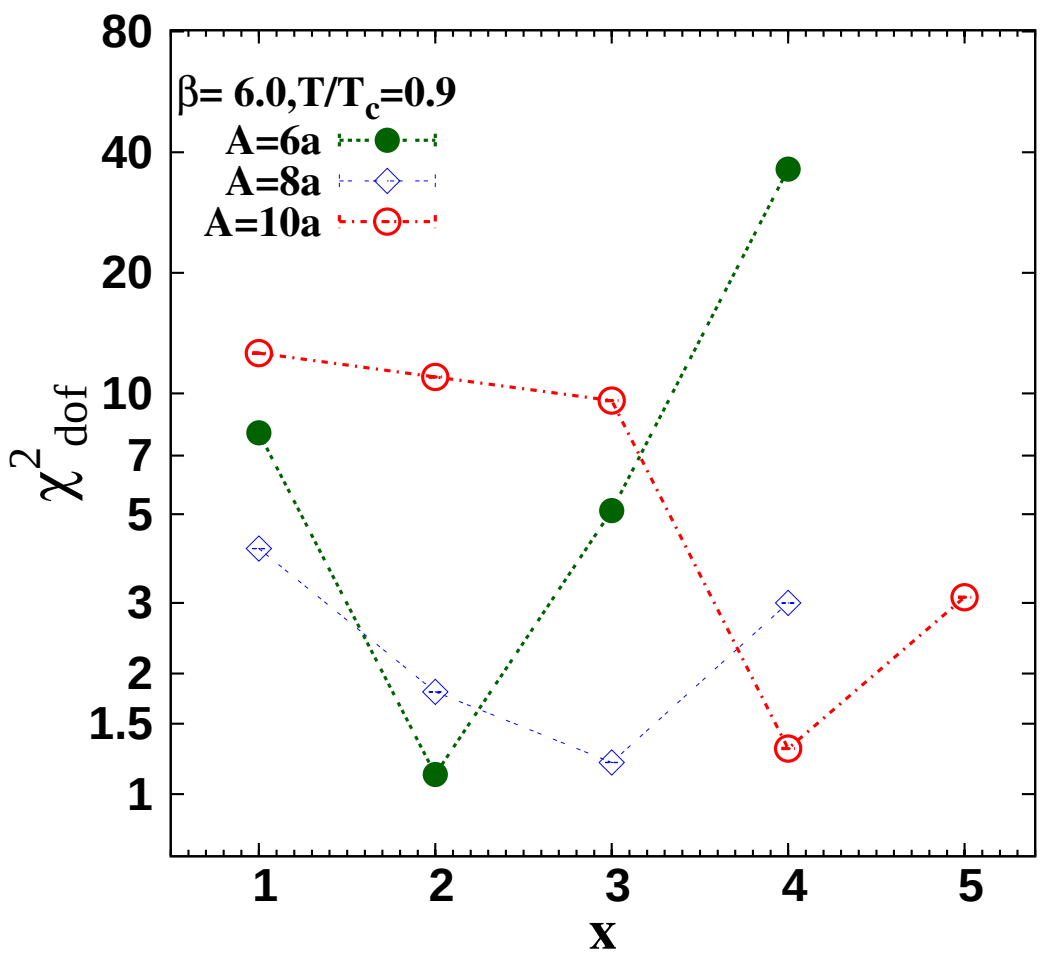

FIG. 18: The returned values of the $\chi^2_{\rm dof}$ for the isosceles base $A = 0.6\,a$, $A = 0.8\,a$ and $A = 1.0\,a$. The plotted values correspond to the fits of the in-plane width of the action density to string model formula Eq. (21) at $T/T_c = 0.9$.

TABLE IV: Similar to Table III: the returned values of the $\chi^2_{\rm dof}$ corresponds to the width in the perpendicular plane $W_z^2$. Fits of the action density to string model formula Eq. (17) at each depicted plane.

(a) $A = 6a$

| Fit range | 5-12 | 6-12 | 7-12 | 8-12 | 9-12 | 10-12 |
|---|---|---|---|---|---|---|
| $\chi^2_{\rm dof}(1)$ | 1.4 | 1.6 | 1.5 | 1.3 | 1.1 | 0.8 |
| $\chi^2_{\rm dof}(2)$ | 2.5 | 2.6 | 1.8 | 1.3 | 1.0 | 0.6 |
| $\chi^2_{\rm dof}(3)$ | 5.7 | 5.6 | 4.6 | 3.4 | 2.3 | 1.4 |
| $\chi^2_{\rm dof}(4)$ | 9.3 | 9.2 | 7.5 | 5.3 | 3.4 | 1.9 |

(b) $A = 8a$

| Fit range | 5-12 | 6-12 | 07-12 | 8-12 | 9-12 | 10-12 |
|---|---|---|---|---|---|---|
| $\chi^2_{\rm dof}(1)$ | 2.1 | 1.2 | 1.4 | 1.4 | 1.5 | 1.3 |
| $\chi^2_{\rm dof}(2)$ | 1.4 | 1.2 | 1.4 | 1.3 | 1.0 | 0.9 |
| $\chi^2_{\rm dof}(3)$ | 1.6 | 1.7 | 1.8 | 1.5 | 1.1 | 0.7 |
| $\chi^2_{\rm dof}(4)$ | 3.7 | 4.4 | 4.0 | 3.0 | 1.9 | 1.0 |

(c) $A = 10a$

| Fit range | 05-12 | 06-12 | 07-12 | 08-12 | 09-12 | 10-12 |
|---|---|---|---|---|---|---|
| $\chi^2_{\rm dof}(1)$ | 14.3 | 2.2 | 1.5 | 1.9 | 2.3 | 2.4 |
| $\chi^2_{\rm dof}(2)$ | 5.7 | 1.4 | 0.9 | 1.2 | 1.4 | 1.5 |
| $\chi^2_{\rm dof}(3)$ | 4.5 | 1.0 | 0.94 | 1.14 | 1.2 | 1.1 |
| $\chi^2_{\rm dof}(4)$ | 2.7 | 1.1 | 1.3 | 1.2 | 1.0 | 0.7 |

immediately before the plane at which $u(x_0) = 0$, i.e, $x_0 - 1$. We shall see in what follows a more clear manifestation of this observation for the analysis at the other temperature $T = 0.9\,T_c$ (just before the deconfinement point).

Table II summarizes the returned $\chi^2_{\rm dof}$ from the fits of perpendicular fluctuations of Eq. 17 to the perpendicular mean-square width profile $W_z^2$listed in Tables VII. Figures 17 plot the corresponding best fits to the string model at the depicted selected planes for each triangle

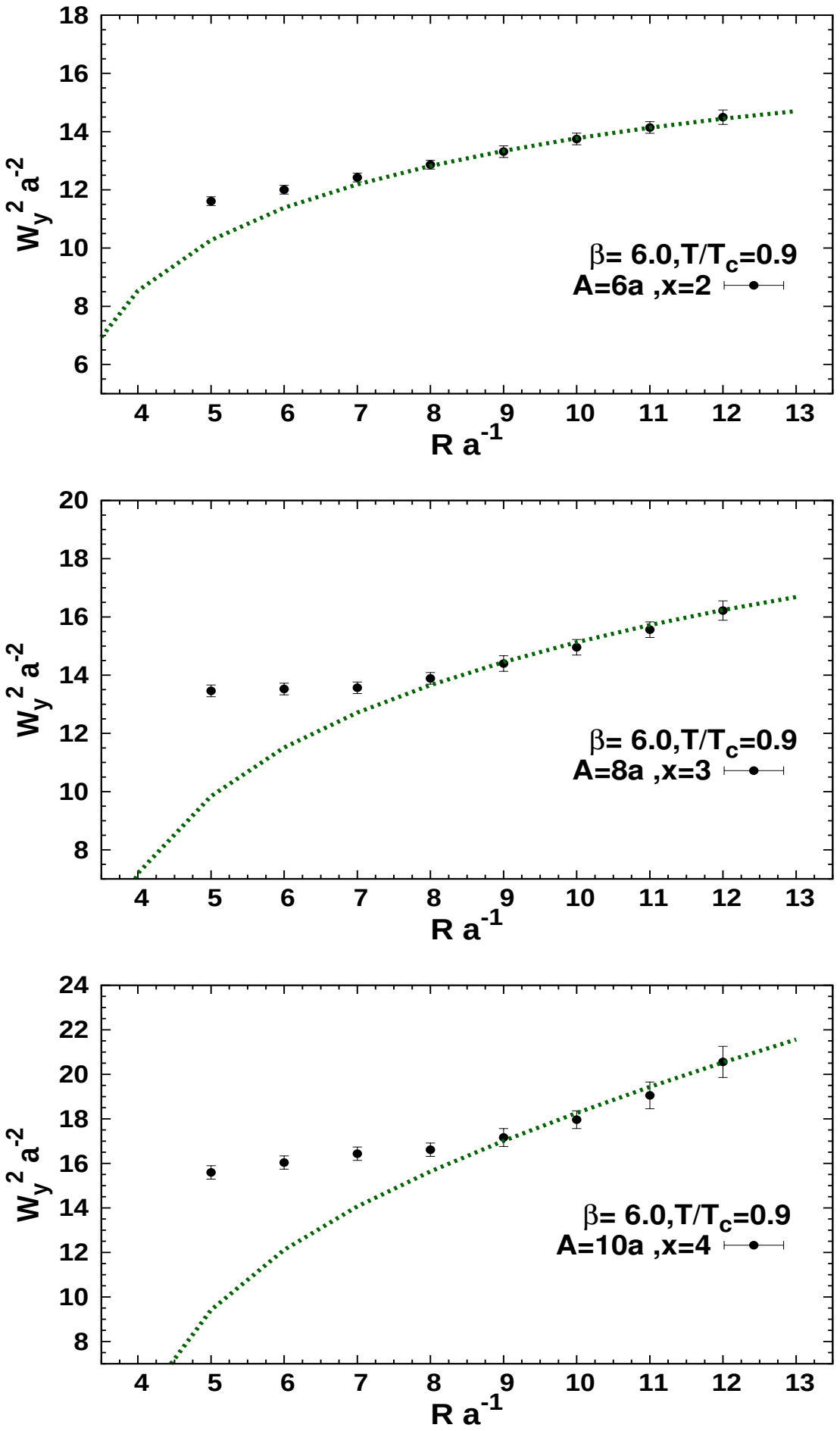

FIG. 19: Similar to Fig. 16 for the width of the in-plane action density $W_y^2$ at temperature $T/T_c = 0.9$ for the depicted transverse planes , $x$.

base parameter.

The perpendicular fluctuations returns good $\chi^2_{\rm dof}$ for widths measured at the plane $x = 1$ and third source $Q_3$ separations commencing from $R > 6$ fm for $A = 6a$. Also similar to the in-plane fluctuations discussed above, good $\chi^2_{\rm dof}$ are obtained for both the plane $x = 1$ and $x = 2$ for the triangular configuration of base lengths $A = 8\,a$ and $A = 10\,a$. Figure 17 demonstrates that for the largest triangle bases the deviations from the string behavior manifest clearly for the width corresponding to third quark separations $R < 6$. In the case of the small base the two strings of the Y-shape are more close in space and self interactions can cause larger deviations to be observed.

At the highest temperature $T/T_c = 0.9$, inspection of Table III shows a similar behavior with respect to the points at small $Q_3$ sources separations with a high value of the returned $\chi^2_{\rm dof}$ if the entire range, i.e.,$R = 5a$ to$R = 12a$ is considered. Good values of $\chi^2_{\rm dof}$ is reached for a third source $Q_3$ separation $R \geq 6$ fm for the isosceles configuration corresponding to base length $A = 6\,a$.

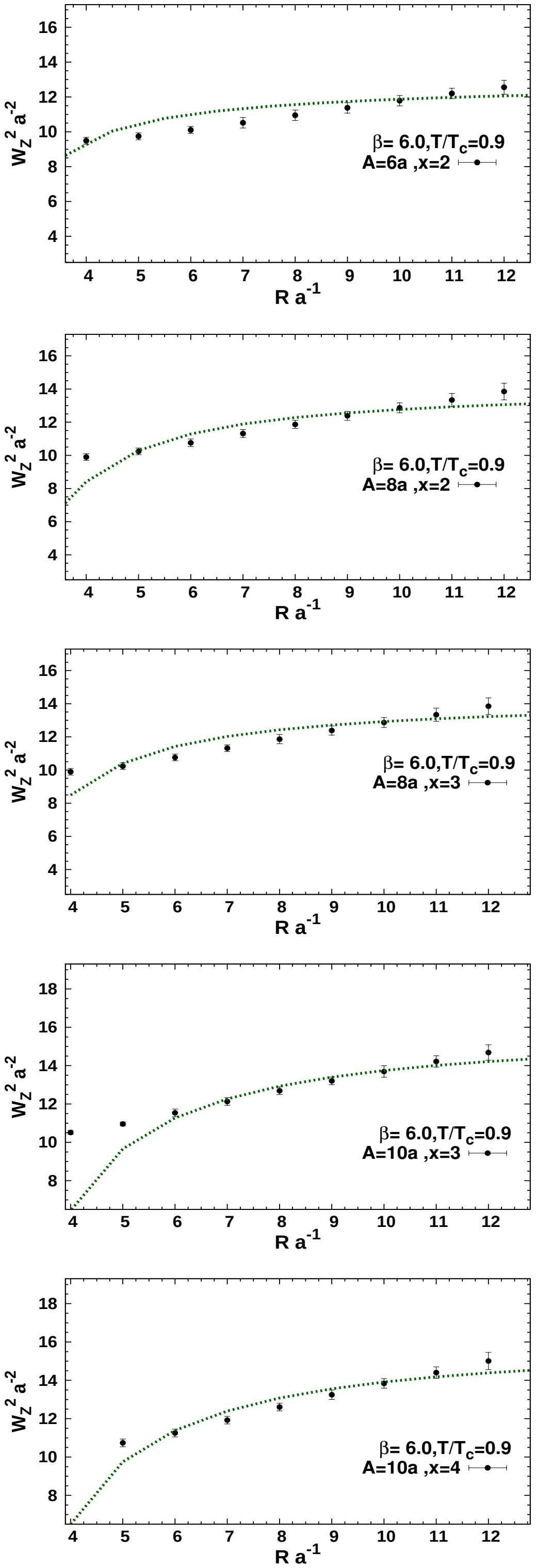

FIG. 20: Similar to Fig. 17 for the width of the perpendicular component of the action density $W_z^2$ at temperature $T/T_c = 0.9$.

This manifests at the plane $x = 2$ which is one plane immediately before the plane $x_0 = 2$ at which $u(x_0) = 0$ i.e, where the two Gaussians coincide as depicted in Fig. 9.

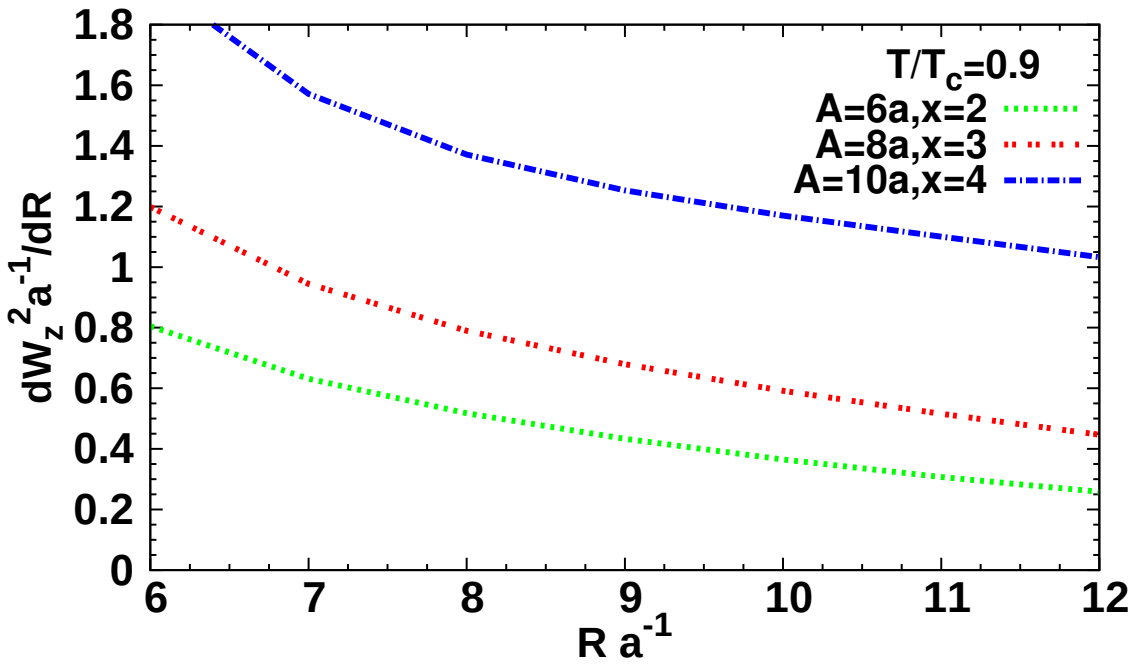

FIG. 21: The first derivative of the in-plane width of the junction with respect to third quark separation according to the best fits returned at the dipected planes and isoscles bases of Fig. 19.

In comparison with the fit behavior at the other temperature $T/T_c = 0.8$, this shows that the best fits change in accord with the change in the position of the transverse plane $x_0$. This is also manifest for both the other two triangular configurations of base length parameters $A = 8\,a$ and $A = 10\,a$ where the best values of $\chi^2_{\rm dof}$ are obtained at the plane $x = 3$and $x = 4$, respectively.

The fits to the two-Gaussian profile and fits to the width of the string fluctuations are two independent sets of profile functions, even though both seem to behave in accord with each other in regard to the change in the position of the plane returning the best fits and that at which the two-Gaussians coincide. This suggests the physical realization that the two Gaussian profiles (Figs. 8 and 9) are a manifestation of the average position due to the fluctuations of the underlying string structure and this point can be conceived as in favor of the Y-string picture.

For completeness, values of $\chi^2_{\rm dof}$ in Table IV have been listed to show the returned fit parameters for the perpendicular fluctuations Eq. IV. Figure IV shows the pattern of width increase at some of the planes returning best fits. These are in general less than its counterpart listed in Table III for the in-plane width, even though we still obtain the best fits at the same planes obtained for the in-plane action-density width.

A general theme of fits discussed above is that, with the points at short distances excluded from the fit, the returned $\chi^2_{\rm dof}$ is smaller. Figures 16 show data points and the corresponding best fits to the string model at the selected planes (which returns the best fits); the Y-string picture at finite temperature poorly describes the lattice data at short distances. The plots depict that the fluctuations of the junction of the Y-free string have a more suppressed profile than the flux tube observed in lattice gauge theory at short distances. Inspection of Fig. 11 shows that the Gaussian-like flux tubes describe a configuration that resembles a $\Delta$ shape at short distances $R = 3\,a, 4\,a, 5\,a$. On the other hand, the growth of the flux-tube diameter, which matches with the predictions of string model, seems to manifest at large source separation where the length of the third string is large enough to reduce the effects of the self-interaction of the third quark $Q_3$ with the junction. This also is consistent with the fact that effective string is a working picture at large source distances [4, 9, 14, 19, 19–21, 21, 22, 49, 60, 61].

The growth pattern, by the movement of one quark sources $Q_3$, in the width of the planes that provide the best fits at the temperature $T/T_c = 0.9$, does not directly indicate a linear growth similar to the more simple mesonic case [21, 49, 60, 61]. This can be read off from the derivative of the string model fluctuations plotted in Fig. 21, of the best fits in Figs.19, with respect to the change of the third quark $Q_3$ position $R$.

From the above discussion, we conclude that the Y-string picture with a minimal length of the string entails a mean-square width of its quantum fluctuations which is consistent with the lattice-gauge data at large color-source separation. The Y-string configurations provide good fits for configurations with minimal length $R$ greater than 0.7 fm for both the in-plane and off-plane width profile. This result shows the stringlike behavior on a distance scale which is consistent with that observed earlier for the mesonic strings.

## V. SUMMARY AND PROSPECTIVE

In this paper we discussed the baryonic stringlike behavior in the profile of the gluonic flux of a three quark system in pure SU(3) Yang-Mills vacuum at finite temperature. The gluon flux is measured as a correlation between the action density operator and three traced (gauge-invariant) Polyakov lines. Measurements of the mean square width have been taken near the end of QCD phase diagram, $T/T_c = 0.8$ and just before the deconfinement point $T/T_c = 0.9$.

For noise reduction, an average over the configuration space has been calculated over 500 independent bins. In each bin an average over 20 measurement has been calculated in addition to average over space-time symmetries. The two lattices analyzed here are cooled with a stout-link smoothing algorithm with a number of sweeps such that the physics of focus is preserved in a systematic and controlled manner [49, 71].

The Y-baryonic string model has been discussed at high temperature for the width profile of the junction. The leading order solution presented in Ref [30] has been considered here. The thermal effects, however, have been incorporated into the formulas accounting for the mean square width of the string's junction.

We have shown a qualitative description of the rendered action density profile in the context of the consistency with the stringlike behavior. The qualitative feature of these density plots at large source separation does not seem to indicate an action density pattern resembling the shape of tubes that would form around the perimeter of the three quarks. The distribution is a "filled" $\Delta$ and displays a peak close to the geometrical center of the triangle. In addition, the transverse profile of the action density fits to a double Gaussian function. The profile of the double-Gaussian describes a clear Y-shaped system of Gaussian flux-tubes. The revealed configurations of these Gaussian flux-tubes is dynamic and reconfigures in accord with the quark configuration and temperature. The total length of the three Gaussian flux-tubes is maximal for the temperature just before the deconfinement point. However, the length approaches the geometrical minimal length at the temperature near the end of the QCD plateau $T/T_c = 0.8$. The flux density shows an asymmetric width profile between that in the plane of the three quark and that in the perpendicular direction which indicates a greater restoring force in the quarks' plane.

The lattice data for the mean-square width of the gluonic action density has been compared to the corresponding width based on string model at finite temperature. We have revealed the characteristics of the growth pattern of the gluonic action-density for three sets of geometrical 3-quark configurations with respect to fits to the Y-string model. The planes close to the junction in profile of the double-Gaussian return good fits to the width of the junction fluctuations in baryonic string model only for large quark separation for both the considered temperatures.

The analysis presented here is of particular relevance to the confining string models, since reports on effects of bosonic strings are usually discussed on the level of the mesonic flux tubes. Apart from the simulations first presented on the Y-string effects of the 3-Potts model [29, 34], it is the our first examination of the effects of the Y-bosonic strings in the action density of the quenched QCD, to the best of our knowledge. The stringlike behavior of the confining flux tubes at the two temperature scales enabled increased insight into the dynamics of the profile with the temperature changes. It would be insightful as well to examine the parametrization ansatz of the confining potential. We report this separately elsewhere.

In the light of the present discussion which focused on revealing the Y-string aspects of the $\Delta$-shaped action density manifesting in the baryon at finite temperature [27, 28], in addition to the observation that the string tension changes only by 10% for the temperature $T/T_c = 0.8$ near the end of the QCD plateau region [72]. These pose the question whether the revealed color map in baryons at the temperatures considered here could be a potential form for the exact geometry of the flux tubes arrangements in the baryon, if the analysis with Polyakov loops is extended to the low temperature regime of pure SU(3) Yang-Mills theory. In addition, it would be interesting to extend the above analysis to reveal the Y-string effects at low temperatures.

We suggest a consistent inclusion of a UV-filtering step into the updating cycles of the Lüscher-Weiss(LW) multi-level algorithm to probe the energy distribution of static baryons in the low temperature regimes of the quenched theory [73]. This technique is expected to contribute to the efficiency of the (LW) algorithm and to reduce the computational time to extend the present analysis to lower temperatures, which is the goal of our next project.

**Acknowledgments**

We would like to thank Dimtri Zanin and Shigemi Ohta for useful comments. This work has been supported by NSFC Grants (Nos. 11035006, 11175215, 11175220), and the Hundred Talent Program of the Chinese Academy of Sciences (Y101020BR0).

## VI. APPENDIX A: STRING PARTITION FUNCTION

We are interested in calculating the string's thickness at the junction position. In the following, we follow the same procedures of the calculations presented in Refs. [29, 30] however, taking into account a convoluted fluctuations $\varphi \to \int_{-\infty}^{\infty} \phi(\tau)\psi(t-\tau)d\tau$ to incorporate thermal effects with the fluctuations as discussed in Sec. II.

The calculation of the corresponding partition function Eq. (9) requires evaluating the integral over the minimal area swept due to perpendicular fluctuations $\varphi$, and the determinant of the Laplacian.

Conformally mapping the string's blade $a$ to a rectangle [29] see Fig. 22

$$f_a(z) = z + \frac{1}{\sqrt{T}} \sum_{w\neq 0} \frac{\boldsymbol{e}_a \cdot \boldsymbol{\varphi}_w \psi(wL_a)}{\sinh(wL_a)} e^{wz}. \quad (28)$$

The minimal area solution for a fixed junction configuration

$$\boldsymbol{\xi}_{min,a} = \frac{1}{\sqrt{T}} \sum_{w} \boldsymbol{\varphi}_{w,\perp a} \psi_w \frac{\sinh(ws)}{\sinh(wL_a)} e^{iwt}, \quad (29)$$

taking into account that the minimal-area solution for a fixed position of the junction, $\boldsymbol{\xi}_{min,a}(t,s)$, is harmonic and satisfies the boundary conditions

$$\triangle \boldsymbol{\xi}_{min,a} = 0\,, \qquad \boldsymbol{\xi}_{min,a}(t, L_a + \boldsymbol{e}_a \cdot \boldsymbol{\varphi}(t)) = \boldsymbol{\varphi}_{\perp a}(t)\,. \quad (30)$$

The integral in Eq. (9) would then read

$$\int_{\Gamma_a} d^2\zeta \sum_i \frac{\partial \boldsymbol{\xi}_{\mathrm{min},a}}{\partial \zeta_i} \cdot \frac{\partial \boldsymbol{\xi}_{\mathrm{min},a}}{\partial \zeta_i} = \sum_w w \coth(wL_a) |\boldsymbol{\varphi}_{w,\perp a}|^2 \psi_w^2.$$

On the other hand, Alvarez-Polyakov formula [12] allows for the calculation of the change in the determinant of the Laplacian with respect to holomorphic mappings $f(z)$ from two domains $\Gamma \to \tilde{\Gamma}$

$$\ln \frac{\det(-\triangle_\Gamma)}{\det(-\triangle_{\tilde{\Gamma}})} = \frac{1}{12\pi} \int_{\partial\Gamma} d\tau \, \frac{\epsilon_{ij} z'^i z''^j}{z'^2} \ln |\partial_z f|^2 + \frac{1}{12\pi} \int_\Gamma d^2 z \, \partial_z \ln |\partial_z f|^2 \partial_{\bar{z}} \ln |\partial_z f|^2\,, \quad (31)$$

with $z(\tau)$ is an arbitrary parametrization of $\partial\Gamma$ and $z' = dz/d\tau$. The determinant in Eq. (9) is obtained by mapping the domain $\Gamma_a$ conformally to a rectangle $L'_a \times T$.

Taking into account the change in the Laplacian

$$\triangle_{\Gamma_a} = e^{2\rho_a(z)} \triangle_{L'_a \times T}\,, \quad \rho_a(z) = -\frac{1}{2} \ln |\partial_z f_a|^2\,. \quad (32)$$

Using the above conformal map Eq. (28), we obtain to leading order

$$\ln \frac{\det(-\triangle_\Gamma)}{\det(-\triangle_{\tilde{\Gamma}})} = \frac{1}{12\pi} \sum_w w^3 |\boldsymbol{e}_a \cdot \boldsymbol{\varphi}_w|^2 \coth(wL_a) \psi(w)\,, \quad (33)$$

Further conformally mapping the above into a circle

$$\det(-\triangle_{L'_a \times T}) = \eta^2 \left( \frac{iT}{2L'_a} \right)\,, \quad (34)$$

where $\eta(\tau)$ denotes the Dedekind $\eta$ function.

Making use Eqs. (32) and (33), the determinant of the Laplacian with respect to the blade $a$ would then read

$$\det(-\triangle_{\Gamma_a}) = \eta^2 \left( \frac{iT}{2L'_a} \right) \times \exp\left( -\frac{1}{12\pi} \sum_w w^3 \coth(wL_a) |\boldsymbol{e}_a \cdot \boldsymbol{\varphi}_w|^2 \psi_w^2 \right)\,. \quad (35)$$

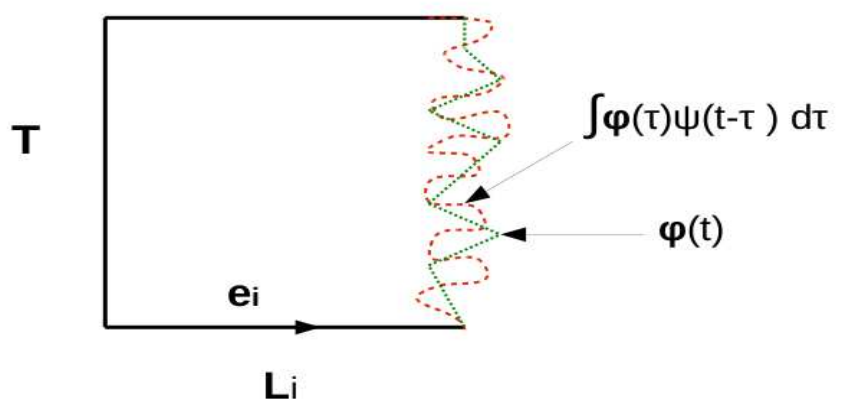

FIG. 22: The domain $\Gamma_a$ is conformally mapped into a rectangle $L'_a \times T$, the first order fluctuations are convoluted with a smoothing scalar $\psi$.

## VII. APPENDIX B: THE JUNCTION'S WIDTH

The thickness of the string at the junction can be calculated taking the expectation value of $\boldsymbol{\varphi}^2$

$$\langle \boldsymbol{\varphi}^2 \rangle = \frac{\int \mathcal{D}\boldsymbol{\varphi}\, \boldsymbol{\varphi}^2 \psi_w^2 e^{-S}}{\int \mathcal{D}\boldsymbol{\varphi}\, e^{-S}} \,. \tag{36}$$

The above second moment of the junction can be decomposed into perpendicular $z$ and parallel (in-plane) $xy$ fluctuations

$$\langle \boldsymbol{\varphi}^2 \rangle = \langle {\boldsymbol{\varphi}_z}^2 \rangle + \langle {\boldsymbol{\varphi}_{xy}}^2 \rangle = \frac{I_{z,2}}{I_{z,0}} + \frac{I_{xy,2}}{I_{xy,0}} \,, \tag{37}$$

where

$$I_{z,2} = \int \mathcal{D}\boldsymbol{\varphi}_z \boldsymbol{\varphi}_z^2 \exp\Bigg\{ -\frac{1}{2}\sum_w \Big[ m w^2 + \sigma w \sum_a \coth(w\, L_a) \Big] \times |\boldsymbol{\varphi}_{w,z}|^2 \Bigg\} \,, \tag{38}$$

$$I_{xy,2} = \int \mathcal{D}\boldsymbol{\varphi}\boldsymbol{\varphi}^2 \exp\Bigg\{ \sum_w \Big[ -\frac{1}{2}\Big( m w^2 + \sigma w \sum_a \coth(w L_a)\Big) \times |\boldsymbol{\varphi}_w|^2 + |\varphi_{w,x}|^2 A_x + |\varphi_{w,y}|^2 A_y + |\varphi_{w,y}|^2 A_y + 2\left((\varphi_{w,x}.\varphi_{w,y})\right) A_{\rm xy} \Big] \Bigg\} \,. \tag{39}$$

with $A_x$, $A_y$ and $A_{xy}$ defined as in Eq. (21). Orthogonalizing the fluctuations for parallel fluctuations the above moments would then read

$$I_{x,2} = \int \mathcal{D}\boldsymbol{\varphi}_x \boldsymbol{\varphi}_x^2 \exp\Bigg\{ \sum_w \Big[ -\frac{1}{2}\Big( F(w) + G_x(w)\Big) |\boldsymbol{\varphi}_{w,x}|^2 \Bigg\} \,.$$

$$I_{y,2} = \int \mathcal{D}\boldsymbol{\varphi}_y \boldsymbol{\varphi}_y^2 \exp\Bigg\{ \sum_w \Big[ -\frac{1}{2}\Big( F(w) + G_y(w)\Big) |\boldsymbol{\varphi}_{w,y}|^2 \Bigg\} \,.$$

$$I_{z,2} = \int \mathcal{D}\boldsymbol{\varphi}_z \boldsymbol{\varphi}_z \exp\Bigg\{ -\frac{1}{2}\sum_w R(w) |\boldsymbol{\varphi}_{w,z}|^2 \Bigg\}.$$

$F(w)$, and $G(w)$ are defined as

$$\begin{aligned} F(w) &= A_{x,w} + A_{y,w} \\ G(w) &= (A_{xy,w}^2 + (A_{x,w} - A_{y,w})^2)^{1/2} \\ R(w) &= m w^2 + \sigma w \sum_i \coth(w L_i)\psi(w, L_i) \end{aligned} \tag{40}$$

Solving for the above Gaussian integrals

$$\begin{aligned} \langle \boldsymbol{\varphi}_x^2 \rangle = \frac{I_{x,2}}{I_{x,0}} &= \frac{2}{T}\sum_{w>0} \frac{1}{F(w) - G(w)} \,, \\ \langle \boldsymbol{\varphi}_y^2 \rangle = \frac{I_{y,2}}{I_{y,0}} &= \frac{2}{T}\sum_{w>0} \frac{1}{F(w) + G(w)} \,, \\ \langle \boldsymbol{\varphi}_z^2 \rangle = \frac{I_{z,2}}{I_{z,0}} &= \frac{2}{T}\sum_{w>0} \frac{1}{R(w)} \,. \end{aligned} \tag{41}$$

With $w = 2\pi n/T$

## VIII. APPENDIX C

TABLE V: The width of the flux-tube $W_y(x)$ at each consecutive transverse plane $x_i$ from the quarks forming the base, $A$, of the isosceles triangle. The measurements for base sources separation distance $A = 6\,a$ for the temperature $T/T_c = 0.8$ are indicated as a function of the third quark position, $Q_3$.

| plane $Q_3 = R/a$ | $x = 1$ | $x = 2$ | $x = 3$ | $x = 4$ | $x = 5$ | $x = 6$ | $x = 7$ | $x = 8$ | $x = 9$ | $x = 10$ | $x = 12$ | $x = 13$ |
|---|---|---|---|---|---|---|---|---|---|---|---|---|
| A=0.6 fm | | | | | | | | | | | | |
| 07 | 9.7(1) | 9.6(0) | 9.3(0) | 9.0(1) | 8.6(1) | 8.2(1) | | | | | | |
| 08 | 10.0(1) | 9.9(1) | 9.7(1) | 9.5(1) | 9.2(1) | 8.7(1) | 8.2(1) | | | | | |
| 09 | 10.2(1) | 10.3(1) | 10.2(1) | 10.0(1) | 9.7(2) | 9.3(2) | 8.8(2) | 8.2(1) | | | | |
| 10 | 10.6(1) | 10.8(2) | 10.6(2) | 10.5(2) | 10.3(2) | 9.9(2) | 9.4(2) | 8.8(2) | 8.0(1) | | | |
| 11 | 10.9(2) | 11.5(3) | 11.2(3) | 11.1(3) | 10.9(3) | 10.6(3) | 10.2(2) | 9.5(2) | 8.6(1) | 7.8(1) | | |
| 12 | 11.3(2) | 12.5(5) | 12.1(4) | 11.8(4) | 11.6(4) | 11.3(4) | 11.0(3) | 10.3(2) | 9.4(2) | 8.4(2) | 7.7(1) | |
| 13 | 11.7(3) | 13.9(7) | 13.2(6) | 12.7(6) | 12.4(6) | 12.3(6) | 12.1(5) | 11.5(3) | 10.4(2) | 9.2(2) | 8.3(2) | 7.9(2) |
| A=0.8 fm | | | | | | | | | | | | |
| 07 | 11.3(4) | 11.2(4) | 11.3(1) | 10.8(1) | 10.2(1) | 9.5(2) | | | | | | |
| 08 | 11.5(4) | 11.4(4) | 11.7(1) | 11.2(1) | 10.6(1) | 10.0(2) | 9.2(2) | | | | | |
| 09 | 11.9(4) | 11.6(4) | 11.7(4) | 11.6(1) | 11.1(2) | 10.5(2) | 9.7(2) | 8.8(2) | | | | |
| 10 | 12.4(5) | 12.1(4) | 11.8(2) | 11.9(2) | 11.4(2) | 10.9(2) | 10.2(2) | 9.3(2) | 8.2(2) | | | |
| 11 | 13.1(5) | 12.5(5) | 12.1(2) | 12.1(2) | 11.5(2) | 11.0(2) | 10.4(1) | 9.6(2) | 8.5(3) | 7.5(3) | | |
| 12 | 13.8(5) | 13.0(5) | 12.4(2) | 12.1(2) | 11.3(2) | 10.8(2) | 10.3(1) | 9.6(2) | 8.6(4) | 7.6(4) | 6.8(4) | |
| A=1.0 fm | | | | | | | | | | | | |
| 07 | 13.4(2) | 12.7(2) | 13.7(1) | 12.9(0) | 12.2(1) | 11.4(2) | | | | | | |
| 08 | 13.6(2) | 12.9(2) | 13.8(2) | 13.0(1) | 12.3(0) | 11.5(1) | 10.6(2) | | | | | |
| 09 | 14.0(3) | 13.3(2) | 13.7(3) | 12.9(2) | 12.1(1) | 11.5(1) | 10.6(2) | 9.5(2) | | | | |
| 10 | 14.5(4) | 13.8(4) | 13.3(4) | 11.7(2) | 11.0(2) | 10.4(2) | 9.4(3) | 8.1(4) | | | | |
| 11 | 15.0(5) | 14.5(6) | 13.9(3) | 10.9(4) | 10.2(4) | 9.6(5) | 9.0(6) | 7.9(6) | 6.6(6) | | | |
| 12 | 15.5(7) | 15.4(8) | 14.8(6) | 9.7(7) | 8.9(7) | 8.3(8) | 7.8(9) | 7(1) | 6(1) | 5(1) | | |

TABLE VI: same as Table V; however the width measurements are taken at the temperature $T/T_c = 0.9$.

| plane $Q_3 = R/a$ | $x = 1$ | $x = 2$ | $x = 3$ | $x = 4$ | $x = 5$ | $x = 6$ | $x = 7$ | $x = 8$ | $x = 9$ | $x = 10$ | $x = 12$ | $x = 13$ | $x = 14$ |
|---|---|---|---|---|---|---|---|---|---|---|---|---|---|
| A=0.6 fm | | | | | | | | | | | | | |
| 07 | 12.9(2) | 12.4(2) | 12.5(0) | 12.0(1) | 11.4(2) | 10.9(3) | | | | | | | |
| 08 | 13.5(1) | 12.9(3) | 12.9(0) | 12.6(1) | 12.1(1) | 11.4(2) | 10.7(3) | | | | | | |
| 09 | 13.6(6) | 13.3(2) | 13.2(0) | 13.0(0) | 12.7(1) | 12.1(2) | 11.4(2) | 10.6(3) | | | | | |
| 10 | 13.9(3) | 13.6(1) | 13.4(1) | 13.4(0) | 13.2(0) | 12.8(1) | 12.2(2) | 11.3(2) | 10.5(3) | | | | |
| 11 | 14.1(2) | 14.1(2) | 14.2(6) | 13.6(1) | 13.6(1) | 13.4(1) | 13.0(1) | 12.2(2) | 11.2(2) | 10.2(3) | | | |
| 12 | 14.3(2) | 14.5(6) | 14.7(2) | 13.7(2) | 13.9(2) | 13.9(1) | 13.7(1) | 13.1(1) | 12.1(2) | 10.9(2) | 9.8(2) | | |
| A=0.8 fm | | | | | | | | | | | | | |
| 07 | 13.2(1) | 13.3(2) | 13.6(1) | 13.5(3) | 13.3(2) | 12.4(3) | | | | | | | |
| 08 | 13.5(1) | 13.7(2) | 13.9(3) | 14.5(0) | 14.0(1) | 13.0(2) | 12.0(3) | | | | | | |
| 09 | 13.9(1) | 14.1(2) | 14.4(3) | 15.4(1) | 14.7(0) | 13.8(1) | 12.7(2) | 11.7(3) | | | | | |
| 10 | 14.1(2) | 14.5(3) | 15.0(4) | 15.8(2) | 15.4(1) | 14.7(1) | 13.7(2) | 12.5(2) | 11.3(3) | | | | |
| 11 | 14.4(2) | 14.9(4) | 15.6(4) | 16.2(3) | 15.9(2) | 15.5(1) | 14.7(1) | 13.5(2) | 12.2(3) | 10.9(3) | | | |
| 12 | 14.6(2) | 15.3(5) | 16.2(7) | (5) | 16.5(3) | 16.2(2) | 15.7(2) | 14.7(2) | 13.4(2) | 11.8(2) | | | |
| A=1.0 fm | | | | | | | | | | | | | |
| 07 | 17.1(2) | 16.7(2) | 16.3 (1) | 16.4(5) | 15.9(1) | 14.5(3) | | | | | | | |
| 08 | 17.6(3) | 17.2(2) | 16.7 (1) | 16.6(2) | 16.7(0) | 15.2(2) | 13.8(3) | | | | | | |
| 09 | 18.0(3) | 17.7(2) | 17.4 (2) | 17.1(3) | 17.5(1) | 16.1(1) | 14.5(2) | 13.0(3) | | | | | |
| 10 | 18.5(4) | 18.3(3) | 18.1 (4) | 17.9(5) | 18.3(2) | 17.1(1) | 15.6(1) | 13.9(2) | 12.4(3) | | | | |
| 11 | 19.0(4) | 19.0(5) | 19.0 (6) | 19.1(5) | 19.1(4) | 18.1(2) | 16.8(1) | 15.1(2) | 13.4(3) | 11.7(3) | | | |
| 12 | 19.3(5) | 19.7(6) | 20 (6) | 20.6(7) | 20.1(5) | 19.2(4) | 18.1(3) | 16.5(2) | 14.7(2) | 12.8(3) | 11.2(4) | | |

TABLE VII: The width of the perpendicular width of the flux-tube to the 3Q plane at each consecutive transverse plane $x_i$ from the quarks forming the base, $A$, of the isosceles triangle. The measurements are taken at the temperature $T/T_c = 0.8$.

| plane $Q_3 = R/a$ | $x=1$ | $x=2$ | $x=3$ | $x=4$ | $x=5$ | $x=6$ | $x=7$ | $x=8$ | $x=9$ | $x=10$ | $x=11$ |
|---|---|---|---|---|---|---|---|---|---|---|---|
| A=0.6 fm | | | | | | | | | | | |
| 04 | 6.7(1) | 6.7(1) | 6.7(1) | | | | | | | | |
| 05 | 6.9(1) | 6.9(1) | 6.9(1) | 6.8(1) | | | | | | | |
| 06 | 7.1(1) | 7.1(1) | 7.1(1) | 7.1(1) | 7.0(1) | | | | | | |
| 07 | 7.3(1) | 7.3(1) | 7.4(0) | 7.4(1) | 7.3(1) | 7.2(1) | | | | | |
| 08 | 7.5(1) | 7.6(1) | 7.7(1) | 7.7(0) | 7.6(1) | 7.5(1) | 7.4(1) | | | | |
| 09 | 7.7(1) | 7.8(1) | 7.9(1) | 8.0(1) | 8.0(1) | 7.9(1) | 7.8(1) | 7.8(2) | | | |
| 10 | 7.9(1) | 8.0(1) | 8.3(1) | 8.4(1) | 8.5(1) | 8.5(1) | 8.4(1) | 8.4(2) | 8.5(2) | | |
| 11 | 8.1(1) | 8.4(1) | 8.7(1) | 9.0(1) | 9.1(1) | 9.1(1) | 9.1(2) | 9.2(2) | 9.3(2) | 9.5(4) | |
| 12 | 8.5(2) | 8.9(2) | 9.3(2) | 9.7(2) | 10.0(2) | 10.1(2) | 10.1(2) | 10.2(2) | 10.4(4) | 10.6(4) | 10.7(5) |
| A=0.8 fm | | | | | | | | | | | |
| 04 | 7.3(1) | 7.2(1) | 7.2(1) | | | | | | | | |
| 05 | 7.5(1) | 7.4(1) | 7.3(1) | 7.3(1) | | | | | | | |
| 06 | 7.6(1) | 7.5(1) | 7.5(1) | 7.4(1) | 7.3(1) | | | | | | |
| 07 | 7.7(1) | 7.7(1) | 7.6(1) | 7.6(1) | 7.5(1) | 7.4(1) | | | | | |
| 08 | 7.9(1) | 7.8(1) | 7.8(1) | 7.8(1) | 7.8(1) | 7.7(2) | 7.6(2) | | | | |
| 09 | 8.0(1) | 8.0(1) | 8.0(1) | 8.0(1) | 8.0(2) | 8.0(2) | 8.0(2) | 8.1(4) | | | |
| 10 | 8.1(2) | 8.2(2) | 8.2(2) | 8.3(2) | 8.3(2) | 8.4(2) | 8.5(2) | 8.7(5) | 9.1(5) | | |
| 11 | 8.3(2) | 8.4(2) | 8.6(2) | 8.7(2) | 8.8(2) | 8.9(2) | 9.2(5) | 9.6(5) | 10.0(8) | 10.3(7) | |
| 12 | 8.6(4) | 8.9(4) | 9.2(2) | 9.4(2) | 9.6(2) | 9.8(5) | 10.1(6) | 10.7(9) | 11.3(7) | 11.6(9) | 11.4(9) |
| A=1.0 fm | | | | | | | | | | | |
| 04 | 7.8(1) | 7.6(1) | 7.5(1) | | | | | | | | |
| 05 | 8.0(1) | 7.8(1) | 7.6(1) | 7.6(1) | | | | | | | |
| 06 | 8.1(1) | 7.9(1) | 7.7(1) | 7.6(1) | 7.5(2) | | | | | | |
| 07 | 8.1(2) | 7.9(1) | 7.7(1) | 7.7(1) | 7.6(2) | 7.5(2) | | | | | |
| 08 | 8.2(2) | 8.0(2) | 7.7(2) | 7.6(2) | 7.6(2) | 7.5(2) | 7.5(4) | | | | |
| 09 | 8.3(2) | 8.0(2) | 7.8(2) | 7.7(2) | 7.6(2) | 7.6(4) | 7.6(4) | 7.8(5) | | | |
| 10 | 8.3(2) | 8.2(2) | 8.0(4) | 7.8(4) | 7.7(4) | 7.7(5) | 7.9(5) | 8.3(6) | 8.7(5) | | |
| 11 | 8.2(5) | 8.4(5) | 8.2(5) | 8.0(5) | 7.9(5) | 7.9(5) | 8.3(8) | 8.9(9) | 9.5(6) | 9.7(4) | |
| 12 | 8.2(8) | 8.6(6) | 8.7(5) | 8.5(5) | 8.3(5) | 8.4(8) | 8.8(8) | 9.6(9) | 10.5(7) | 10.7(4) | 10.3(4) |

TABLE VIII: The percentage difference $e$ between the in-plane width measurements $W_y^{(1)}(x))^2$ with a single Gaussian form [28] relative to the width measured using a two Gaussian form $(W_y^2(x))^2$ calculated as, $e = |((W_y^{(1)}(x))^2 - (W_y^{(2)}(x))^2)/(W_y^{(2)}(x))^2|$, at temperature $T/T_c = 0.8$ and for the selected quark configurations.

| plane $Q_3 = R/a$ | $R=5$ | $R=7$ | $R=9$ | $R=11$ |
|---|---|---|---|---|
| A=0.8 fm | | | | |
| x=1 | 8% | 7% | 7% | 4% |
| x=2 | 5% | 6% | 9% | 9% |
| A=1.0 fm | | | | |
| x=1 | 15.0% | 10% | 8.2% | |
| x=2 | 15.0% | 9% | 2.7% | |

TABLE IX: Same as Table VII; however the perpendicular width of the action density $W_z^2(x)$ has been measured at the temperature $T/T_c = 0.9$.

| plane $Q_3 = R/a$ | $x=1$ | $x=2$ | $x=3$ | $x=4$ | $x=5$ | $x=6$ | $x=7$ | $x=8$ | $x=9$ | $x=10$ | $x=11$ |
|---|---|---|---|---|---|---|---|---|---|---|---|
| A=0.6 fm | | | | | | | | | | | |
| 04 | 9.5(2) | 9.5(2) | 9.5(2) | | | | | | | | |
| 05 | 9.7(2) | 9.7(2) | 9.7(2) | 9.7(2) | | | | | | | |
| 06 | 10.0(2) | 10.1(2) | 10.2(2) | 10.1(2) | 9.9(2) | | | | | | |
| 07 | 10.3(2) | 10.5(2) | 10.7(2) | 10.6(2) | 10.4(2) | 10.2(2) | | | | | |
| 08 | 10.7(2) | 10.9(2) | 11.2(2) | 11.3(2) | 11.2(2) | 10.9(2) | 10.6(4) | | | | |
| 09 | 11.0(2) | 11.4(2) | 11.7(2) | 11.9(2) | 11.9(2) | 11.7(2) | 11.3(4) | 11.0(4) | | | |
| 10 | 11.4(2) | 11.8(2) | 12.2(2) | 12.5(2) | 12.6(2) | 12.6(2) | 12.2(2) | 11.8(4) | 11.3(4) | | |
| 11 | 11.8(4) | 12.2(2) | 12.7(2) | 13.1(2) | 13.3(2) | 13.4(2) | 13.2(4) | 12.8(4) | 12.3(4) | 11.8(5) | |
| 12 | 12.3(4) | 12.7(4) | 13.1(2) | 13.6(2) | 14.0(2) | 14.2(4) | 14.3(4) | 14.0(4) | 13.5(5) | 12.9(5) | 12.5(5) |
| A=0.8 fm | | | | | | | | | | | |
| 04 | 10.1(2) | 10.0(2) | 9.9(2) | | | | | | | | |
| 05 | 10.5(2) | 10.4(2) | 10.2(2) | 10.1(2) | | | | | | | |
| 06 | 10.8(2) | 10.8(2) | 10.8(2) | 10.6(2) | 10.4(2) | | | | | | |
| 07 | 11.2(2) | 11.3(2) | 11.3(2) | 11.2(2) | 11.0(2) | 10.7(4) | | | | | |
| 08 | 11.5(2) | 11.7(2) | 11.9(2) | 11.9(2) | 11.7(2) | 11.4(4) | 11.1(4) | | | | |
| 09 | 11.9(2) | 12.1(2) | 12.4(2) | 12.5(2) | 12.5(2) | 12.2(2) | 11.8(4) | 11.4(4) | | | |
| 10 | 12.3(2) | 12.6(2) | 12.9(2) | 13.1(2) | 13.2(2) | 13.1(2) | 12.7(4) | 12.2(4) | 11.8(5) | | |
| 11 | 12.8(4) | 13.0(2) | 13.3(2) | 13.6(2) | 13.9(2) | 13.9(2) | 13.7(4) | 13.2(4) | 12.7(5) | 12.3(5) | |
| 12 | 13.4(4) | 13.6(4) | 13.8(2) | 14.2(2) | 14.5(2) | 14.7(4) | 14.7(4) | 14.3(5) | 13.8(5) | 13.3(5) | 13.1(5) |
| A=1.0 fm | | | | | | | | | | | |
| 04 | 11.0(2) | 10.7(2) | 10.5(2) | | | | | | | | |
| 05 | 11.4(2) | 11.2(2) | 11.0(2) | 10.7(2) | | | | | | | |
| 06 | 11.8(2) | 11.7(2) | 11.5(2) | 11.3(2) | 11.0(4) | | | | | | |
| 07 | 12.2(2) | 12.2(2) | 12.1(2) | 11.9(2) | 11.6(4) | 11.3(4) | | | | | |
| 08 | 12.6(2) | 12.6(2) | 12.7(2) | 12.6(2) | 12.4(4) | 12.0(4) | 11.7(5) | | | | |
| 09 | 13.0(2) | 13.1(2) | 13.2(2) | 13.2(2) | 13.1(2) | 12.8(4) | 12.4(4) | 12.0(5) | | | |
| 10 | 13.4(2) | 13.5(2) | 13.7(2) | 13.8(2) | 13.9(2) | 13.7(4) | 13.3(4) | 12.8(5) | 12.4(5) | | |
| 11 | 14.1(2) | 14.1(2) | 14.2(2) | 14.4(2) | 14.6(2) | 14.6(4) | 14.3(4) | 13.8(5) | 13.2(5) | 13.0(5) | |
| 12 | 15.0(4) | 14.9(2) | 14.9(2) | 15.0(2) | 15.2(2) | 15.4(4) | 15.3(5) | 14.8(5) | 14.3(5) | 13.9(6) | 13.9(8) |

TABLE X: Same as Table VIII; however the percentage difference has been calculated for in-plane width at the temperature $T/T_c = 0.9$.

| plane $Q_3 = R/a$ | $R=5$ | $R=7$ | $R=9$ | $R=11$ |
|---|---|---|---|---|
| A=0.6 fm | | | | |
| x=1 | 2.3% | 2.0% | 1.0% | 3.0 % |
| x=2 | 5.1% | 3.2% | 0.1% | 5.0% |
| A=0.8 fm | | | | |
| x=1 | 25.2% | 25.7% | 24.4% | 28.4% |
| x=2 | 18.9% | 18.9% | 15.6% | 12.0% |
| x=3 | 4.4 % | 11.0% | 10.4% | 5.1% |
| A=1.0 fm | | | | |
| x=1 | 24.8% | 25.1% | 22.2 % | 23% |
| x=2 | 19.2% | 20.3% | 16.9 % | 15.7% |
| x=3 | 9.0% | 15.3% | 13.2% | 5% |
| x=4 | 1.0% | 6.0% | 5.8 % | 4% |